\documentclass[aip, reprint, longbibliography]{revtex4-1}
\usepackage[utf8]{inputenc}
\usepackage[lining,semibold]{libertine} % a bit lighter than Times--no osf in math
\usepackage[libertine, cmintegrals, bigdelims, vvarbb]{newtxmath}
\usepackage{amsmath}
\usepackage{amsfonts}
\usepackage{mathrsfs}
\usepackage{gensymb}
\usepackage{bbm}
\usepackage{dsfont}

\usepackage{chemformula}
\usepackage[caption=false]{subfig}
\usepackage{chemfig}
\usepackage[version=3]{mhchem}

\usepackage{tikz}
\usetikzlibrary{matrix,positioning,decorations.pathreplacing}

\usepackage{braket}

\usepackage{soul}
\usepackage{xcolor}

\usepackage{scalerel} % To change size things in math environment \scaleto{...}{1pt}
\usepackage{comment}

\usepackage{blkarray}

\usetikzlibrary{arrows, automata}
\usepackage{algorithm}
\usepackage{algpseudocode}

\definecolor{webgreen}{rgb}{0,.5,0}
\definecolor{webbrown}{rgb}{.6,0,0}
\definecolor{grigio}{rgb}{.85,.85,.85} 
\definecolor{RoyalBlue}{rgb}{0.0, 0.14, 0.4}
\definecolor{skyblue1}{rgb}{0.45,0.62,0.81}
\definecolor{skyblue2}{rgb}{0.2,0.39,0.64}
\definecolor{skyblue3}{rgb}{0.13,0.29,0.53}
\definecolor{scarlet1}{rgb}{0.93,0.16,0.16}
\definecolor{scarlet2}{rgb}{0.8,0,0}
\definecolor{scarlet3}{rgb}{0.64,0,0}

\definecolor{g}{gray}{0.50}

\usepackage{hyperref}
\hypersetup{%
    colorlinks=true, linktocpage=true, pdfstartpage=1, pdfstartview=FitV,%
    breaklinks=true, pdfpagemode=UseNone, pageanchor=true, pdfpagemode=UseOutlines,%
    plainpages=false, bookmarksnumbered, bookmarksopen=true, bookmarksopenlevel=1,%
    hypertexnames=true, pdfhighlight=/O,%nesting=true,%frenchlinks,%
    urlcolor=webbrown, linkcolor=RoyalBlue, citecolor=webgreen, %pagecolor=RoyalBlue,%
    pdftitle={},%
    pdfauthor={Benedikt Remlein},%
    pdfsubject={},%
    pdfkeywords={},%
    pdfcreator={pdfLaTeX},%
    pdfproducer={LaTeX REVTeX}%
}

\usepackage[capitalise]{cleveref}

\usepackage{verbatim}

\newcommand{\col}{\color{black}}

\makeatletter
\def\maketag@@@#1{\hbox{\m@th\normalfont\normalsize#1}}
\makeatother

\DeclareMathAlphabet{\mathpzc}{OT1}{pzc}{m}{it}

\begin{document}
	
	\preprint{AIP/123-QED}
	
    \title{What is a chemostat?\\ 
    Insights from hybrid dynamics and stochastic thermodynamics}

	\author{Benedikt Remlein}
	\email{benedikt.remlein@uni.lu}
    \affiliation{ 
	Complex Systems and Statistical Mechanics, Department of Physics and Materials Science,
University of Luxembourg, 30 Avenue des Hauts-Fourneaux, L-4362 Esch-sur-Alzette, Luxembourg
}%

	\author{Massimiliano Esposito}%
	\email{massimiliano.esposito@uni.lu}
    \affiliation{ 
		Complex Systems and Statistical Mechanics, Department of Physics and Materials Science,
University of Luxembourg, 30 Avenue des Hauts-Fourneaux, L-4362 Esch-sur-Alzette, Luxembourg
	}%

	\author{Francesco Avanzini}%
	\email{francesco.avanzini@unipd.it}
	\affiliation{ 
	Department of Chemical Sciences, University of Padova, Via F. Marzolo, 1, I-35131 Padova, Italy
}%

	\date{\today}
    
	\begin{abstract}
    At the microscopic scale, 
    open chemical reaction networks are described 
    by stochastic reactions that 
    follow mass-action kinetics
    and are coupled to chemostats.
    % -
    % -
    We show that closed chemical reaction networks
    \---
    with specific stoichiometries imposed by mass-action kinetics
    \---
    behave like open ones
    in the limit where the abundances of a subset of species become macroscopic, 
    thus playing the role of chemostats.
    % -
    We prove that this limit is thermodynamically consistent
    by 
    recovering the 
    local detailed balance condition of open chemical reaction networks
    and 
    deriving the 
    proper expression of the entropy production rate.
    % -
    % -
    In particular, the entropy production rate features two contributions:
    % -
    one accounting for the dissipation of the stochastic reactions,
    the other for the dissipation of continuous reactions 
    controlling the chemostats.
    % -
    % -
    Finally, we illustrate our results for two prototypical examples.
	\end{abstract}
	
	\maketitle 
	
%%%%%%%%%%%%%%%%%%%%%%%%%%%%%%%%%%%%%%%%%%%%%%%%%%%%%%%%%%%%
%%%%%%%%%%%%%%%%%%%%%%%%%%%%%%%%%%%%%%%%%%%%%%%%%%%%%%%%%%%%
%%%%%%%%%%%%%%%%%%%%%%%%%%%%%%%%%%%%%%%%%%%%%%%%%%%%%%%%%%%%
	
	\section{Introduction}
    
    Many complex processes in biosystems and synthetic chemistry arise
    from underlying open chemical reaction networks (CRNs)
    maintained out of equilibrium.~\cite{Yang2021,Hermans2017} 
    % -
    % -
    The energetic cost that enables CRNs to stay out of equilibrium is harnessed from 
    the environment via exchanges of chemical species.~\cite{espo20}
    % - 
    High-energy species are harvested and 
    converted into low-energy species released back into the environment
    to power reactions that would otherwise equilibrate.~\cite{Wachtel2022,bila25}
    % -
    % -
    For instance, in biosystems, 
    cellular metabolism converts glucose into carbon dioxide 
    to continuously produce energetic mediator molecules,
    e.g., adenosine triphosphate.~\cite{voet2010}
    % -
    Similarly, in synthetic chemistry,
    different high-energy-species-to-low-energy-species reactions 
    are used to, for instance,
    produce a directional motion in artificial molecular motors
    or 
    form energetically unfavorable assemblies.~\cite{Munana2018aa, prins2021}
    % -
    
    % -
    In recent years,
    rigorous theoretical frameworks have been developed
    to explain such complex processes, 
    often modeling the environment in terms of chemostats,
    i.e.,
    infinitely large reservoirs of chemical species 
    that remain unaffected by chemical reactions,
    but can evolve as a result of some (externally imposed) driving protocol.
    % -
    % -
    From a dynamical standpoint,
    these theories have been used, for instance,
    to prove the emergence of oscillations and chaos.~\cite{andr08b,Gaspard2020}
    % -
    Furthermore, driving the chemostats has been shown to (potentially) lead to
    a highly nonlinear response of the reaction currents,~\cite{Altaner2015, falasco2019ndr, forastiere2020}
    but it cannot, in general, invert them.~\cite{Marehalli2023}
    % -
    % -
    Correspondingly, thermodynamic theories
    have been derived to 
    quantify the energetic cost of maintaining CRNs out of equilibrium
    independently of whether reactions are described as
    stochastic~\cite{gasp04, schm06, rao18} (microscopic scale)
    or deterministic~\cite{Qian2005, rao16, avan21, avan24} (macroscopic scale) events.
    % -
    These theories have been used to quantify 
    the cost of maintaining coherent oscillations,~\cite{ober22,reml22}
    of powering chemical growth,~\cite{mare24a, mare24b}
    as well as to study the efficiency of central metabolism.~\cite{Wachtel2022, Voor2024}
    % -
    They also define thermodynamic speed limits.~\cite{Yoshimura2021a, Yoshimura2021b} 
    % -
    % -
    
    % -
    While the notion of chemostats has proven invaluable in these frameworks,
    recent studies have begun exploring ways to move beyond it.
    % -
    For instance, 
    the emergence of oscillations in a specific CRN coupled to finite chemostats 
    has been investigated at the microscopic scale,~\cite{frit20}
    while 
    the thermodynamics of CRNs directly coupled to specific exchange processes, 
    instead of chemostats,
    has been studied at the macroscopic scale.~\cite{blokhuis2018, avan22}
    % -
    % -
    However,
    a rigorous explanation of the emergence of chemostats is still missing.
    % -
    In this paper, we bridge the gap between
    a closed description of CRNs, where all species have finite abundances, 
    and an open one, where abundant species are treated as chemostats.
    % -
    % -
    
    % -
    We focus on CRNs at the microscopic scale, 
    where
    species abundances are quantified by molecule numbers
    and
    reactions are stochastic events following mass-action kinetics.
    % -
    % -
    We prove that the dynamics and thermodynamics of open CRNs 
    (summarized in Sec.~\ref{sec:openCRN})
    emerge
    from a partial macroscopic limit 
    of closed CRNs with specific stoichiometries.
    % -

    % -
    To do so, 
    we use a large volume limit~\cite{ball06,menz12,crud12,kang13,ande15,wink17,huft19}
    developed for hybrid CRNs (summarized in Sec.~\ref{sec:hybridDynamics})
    where high-abundant species,
    whose molecule numbers scale (linearly) with the volume,
    are mixed together with low-abundant species,
    whose molecule numbers do not scale with the volume.
    % -
    Crucially, this volume limit is based on
    the Wentzel–Kramers–Brillouin (WKB) 
    ansatz,~\cite{risken,touc09,bres14,touc18,qian21,fala25}
    avoiding physical inconsistencies found in
    previous scalings.~\cite{horo15,Ceccato2018}
    % -
    It leads to a hybrid dynamics.
    % -
    The low-abundant species are 
    interconverted by reactions (called discrete reactions)
    still described in terms of stochastic events.
    % -
    The high-abundant species can be 
    interconverted only by reactions (called continuous reactions) 
    described {\col in terms of deterministic rate equations}.
    % -

    % -
    First, we show that the partial macroscopic limit 
    can only be applied to CRNs
    satisfying specific stoichiometric conditions 
    (in Eqs.~\eqref{eq:ConstraintDiscrete} and~\eqref{eq:ConstraintContinuous}) 
    imposed by mass-action kinetics (Sec.~\ref{sec:HDele}).
    % -
    % -
    When these conditions are met,
    the high-abundant species become (in the large volume limit) chemostats.
    % -
    If there are no continuous reactions,
    the resulting chemostats are autonomous 
    with constant concentrations
    (Subs.~\ref{sec:HybridChemostat}).
    % -
    Otherwise,
    continuous reactions impose a deterministic evolution
    independent of the low-abundant species on
    chemostat concentrations (Subs.~\ref{sec:HybridEvolve}),
    thus physically representing the (externally imposed) driving protocols 
    used in the modeling of open CRNs. 
    % -

    % -
    Second, we prove that the partial macroscopic limit 
    is thermodynamically consistent (Sec.~\ref{sec:TD}).
    % -
    On the one hand, 
    it leads to the local detailed balance condition of open CRNs 
    (Subs.~\ref{sec:HDldb}).
    % -
    On the other hand,
    we derive the corresponding entropy production rate
    (Sec.~\ref{sec:HDepr})
    which results from the sum of two contributions.
    One contribution
    accounts for the dissipation of the discrete reactions
    corresponding to the entropy production rate of open CRNs 
    and does not scale with the volume (Subs.~\ref{subsub:EPRdiscrete}).
    % -
    The other accounts for the dissipation of the continuous reactions
    and scales (linearly) with the volume (Subs.~\ref{sec:EPRcontinuous}).
    % -
    In particular, the latter contribution physically quantifies the dissipation
    of the (externally imposed) protocols that manipulate chemostats 
    and
    is always neglected in the modeling of open CRNs.
    % -

    % -
    Finally, we illustrate our results for two prototypical examples 
    (Sec.~\ref{sec:examples}),
    one featuring a CRN without continuous reactions,
    while the other featuring a CRN with a continuous reaction.
    % -

    %- 
    We summarize and discuss our results in Sec.~\ref{sec:Discussion},
    where we also examine future perspectives.

%%%%%%%%%%%%%%%%%%%%%%%%%%%%%%%%%%%%%%%%%%%%%%%%%%%%%%%%%%%%
%%%%%%%%%%%%%%%%%%%%%%%%%%%%%%%%%%%%%%%%%%%%%%%%%%%%%%%%%%%%
%%%%%%%%%%%%%%%%%%%%%%%%%%%%%%%%%%%%%%%%%%%%%%%%%%%%%%%%%%%%
% \newpage	
	\section{Chemical reaction networks}
	\label{sec:openCRN}

    Chemical reaction networks (CRNs) describe 
    mixtures of reacting chemical species (labeled~$\alpha$) 
    and an abundant non-reacting species called the solvent (labeled~$s$).
    % -
    The solvent plays the role of a thermal and volume reservoir, 
    maintaining the mixture temperature $T$ and volume $\Omega$ constant.
    % -
    % -
    CRNs are said to be \textit{open} 
    when the abundances of some reacting species (labeled~$c$) 
    are controlled
    by unspecified external processes
    and thus act as \textit{chemostats}.
    The other reacting species (labeled~$i$) are called \textit{internal} species.
    % -
    In the absence of chemostats, CRNs are said to be \textit{closed}.
    % - 
    % -
    In the following, 
    we use $n_\alpha$ for the molecule number of species $\alpha$ and 
    $n_s$ for the molecule number of the solvent.
    % -
    The vectors 
    $\boldsymbol n_i \equiv ( \ldots\,, n_i\,, \ldots)$ and 
    $[\boldsymbol c] \equiv ( \ldots\,, [c]\,, \ldots)$ %with $[c] = n_c / \Omega$
    specify the 
    molecule numbers of the internal species
    and the concentrations of the chemostats, respectively.
    % -
    Note that the concentrations $[\boldsymbol c]_t$ can evolve in time
    as a result of some driving protocol imposed by the external processes. 
    In this case, chemostats are said to be \textit{evolving}.
    % -
    Otherwise, 
    if the concentrations $[\boldsymbol c]_t$ are maintained constant 
    by the external processes,
    chemostats are said to be \textit{autonomous}.
    % -
    % -

    % -
    Reactions 
    (labeled $\rho \in \mathcal R \equiv \{ \pm 1\,, \pm 2\,, \ldots\,, n_r\}$)
    interconvert the reacting species and
    are described by the following chemical equation
    \begin{equation}\small
        \sum_i \nu_{i,\rho} \, Z_i + \sum_c \nu_{c,\rho} \, Z_c 
        \ch{<=>[$\rho$][$-\rho$]} 
        \sum_i \, \nu_{i,-\rho} Z_i +\sum_c \, \nu_{c,-\rho} Z_c  \,,
        \label{eq:CRNopen}
    \end{equation}
    where 
    $Z_\alpha$ is the chemical symbol of species $\alpha$, while
    $\nu_{\alpha,\rho}$ denotes its  stoichiometric coefficient in reaction $\rho$. 
    % -
    The net variation of molecule number of species~$\alpha$ in reaction~$\rho$ 
    is specified by
    \begin{equation}
    S_{\alpha,\rho} \equiv \nu_{\alpha,-\rho} -\nu_{\alpha,\rho} \,,
    \end{equation}
    and we define the vectors
    $\mathbf S_{i,\rho}= ( \ldots\,, S_{i,\rho}\,, \ldots)$ and
    $\mathbf S_{c,\rho}= ( \ldots\,, S_{c,\rho}\,, \ldots)$
    for the internal species and the chemostats, respectively.
    % -
    Note that reactions are assumed here to be reversible:
    for every forward reaction $\rho \in \mathcal R$,
    reaction $-\rho \in \mathcal R$ and denotes its backward counterpart. 
    % -

    % -
    The probability $P_t(\boldsymbol n_i)$ of there being $\boldsymbol n_i$ molecules at time~$t$ 
    follows the chemical master equation (CME),
    \begin{equation}\small
        \begin{split}
            \partial_t P_t(\boldsymbol n_i) 
            = \sum\limits_{\rho \in \mathcal R} 
            \big\{
            &R_\rho(\boldsymbol{n}_i-\mathbf S_{i,\rho}|[\boldsymbol{c}]_t)
            P_t(\boldsymbol{n}_i-\mathbf S_{i,\rho})\\ 
            &- R_\rho(\boldsymbol{n}_i|[\boldsymbol{c}]_t)P_t(\boldsymbol{n}_i)
            \big\}
            \,,
        \end{split}
        \label{eq:CME}
    \end{equation}
    where the reaction rates are assumed to satisfy mass-action kinetics,~\cite{gasp04}
    \begin{equation}    
        R_\rho(\boldsymbol{n}_i|[\boldsymbol{c}]_t) 
        \equiv 
        \Omega k_\rho 
        \prod_i \frac{n_i!}{(n_i-\nu_{i,\rho})!\,\Omega^{\nu_{i,\rho}}}
        \prod_c [c]_t^{\nu_{c,\rho}} \,,
        \label{eq:MassActionOpen}
    \end{equation} 
    with $k_{\rho}>0$ being the kinetic rate constant of reaction $\rho$.
    The volume dependence in Eq.~\eqref{eq:MassActionOpen} guarantees 
    the correct correspondence between stochastic and deterministic reaction rates
    in the macroscopic limit 
    (i.e., for large molecule numbers of all species $\{n_\alpha\}$ 
    and a large volume $\Omega$, 
    but
    with finite concentrations $\{n_\alpha/\Omega\}$).~\cite{Kurtz1972}
    % -
    % -
    Furthermore, 
    we stress that mass-action kinetics is satisfied for 
    elementary reactions in ideal dilute mixtures, 
    namely, species are noninteracting 
    and the solvent is much more abundant than all the other species. 
    If reactions are not elementary or species interact, 
    reaction rates might not satisfy mass-action kinetics.
    In this work, we consider only elementary reactions in ideal dilute mixtures.
    % -
    % -
    Finally, we stress that in the following we will use
    $R_\rho(\boldsymbol{n}_i)$ (resp. $R_\rho([\boldsymbol{c}]_t)$)
    for the rate of a reaction 
    not involving (resp. involving only) chemostats,
    instead of the general symbol $R_\rho(\boldsymbol{n}_i|[\boldsymbol{c}]_t)$.
    % -
    
    % -
    Thermodynamic consistency imposes that reaction rates~\eqref{eq:MassActionOpen} 
    are related to the Gibbs free energy 
    via the so-called local detailed balance condition,~\cite{rao18}
    \begin{equation}
        \ln \frac
        {R_{\rho}(\boldsymbol{n}_i|[\boldsymbol{c}]_t)
        }{
        R_{-\rho}(\boldsymbol{n}_i+\mathbf S_{i,\rho}|[\boldsymbol{c}]_t)} 
        = 
        - \big\{
        \Delta_\rho g(\boldsymbol{n}_i) + 
        \boldsymbol{\mu}([\boldsymbol{c}]_t)\cdot \mathbf S_{c,\rho}
        \big\}
        \,.
        \label{eq:LDBopen}
    \end{equation}
    % -
    Here, 
    $\Delta_\rho g(\boldsymbol{n}_i)  \equiv 
    g(\boldsymbol{n}_i+\mathbf S_{i,\rho}) 
    - g(\boldsymbol n_i) $ 
    is the variation of the Gibbs free energy along reaction $\rho$
    due to the internal species
    and
    \begin{equation}
        g(\boldsymbol{n}_i) \equiv 
        \sum_i 
        \big\{
        ({\mu}_i^\circ - \ln n_s) {n}_i + 
        \ln{n}_i!
        \big\}\,,
        \label{eq:GibbsFreeEnergy}
    \end{equation}
    is the Gibbs free energy 
    (given in units of temperature $T$ times Boltzmann constant $k_{\mathrm B}$)
    of ideal dilute mixtures with~$\boldsymbol{n}_i$ molecules of internal species,
    with  
    ${\mu}_i^\circ$ being the standard chemical potential of species~$i$.
    % -
    On the other hand,
    ${\boldsymbol{\mu}([\boldsymbol{c}]_t)\cdot \mathbf S_{c,\rho}}$
    is the variation of the Gibbs free energy along reaction $\rho$
    due to the chemostats
    and
    \begin{equation}
        \boldsymbol{\mu}([\boldsymbol{c}]_t) 
        \equiv 
        \boldsymbol{\mu}_c^\circ 
        + 
        \ln([\boldsymbol{c}]_t/[s])
        \,,
        \label{eq:ChemicalPotential}
    \end{equation}
    is the vector of their chemical potentials,
    with 
    $\boldsymbol{\mu}_c^\circ = (\ldots\,, {\mu}_c^\circ\,, \ldots)$ being
    their standard chemical potentials,
    $\ln[\boldsymbol{c}] = (\ldots\,, \ln[{c}]\,, \ldots)$,
    and $[s] = n_s / \Omega$.
    % -

    % -
    The local detailed balance condition~(\ref{eq:LDBopen}) ensures 
    that the (average) entropy production rate,
    quantifying the average dissipated free energy,
    is specified by~\cite{gasp04,schm06,rao16,rao18}
    \begin{equation}\small
        \braket{\dot \Sigma}_t = 
        \sum\limits_{\rho,\boldsymbol{n}_i} 
        R_{\rho}(\boldsymbol{n}_i|[\boldsymbol{c}]_t)P_t(\boldsymbol{n}_i)
        \ln \frac{
        R_{\rho}(\boldsymbol{n}_i|[\boldsymbol{c}]_t)P_t(\boldsymbol{n}_i)
        }{
        R_{-\rho}(\boldsymbol{n}_i+\mathbf S_{i,\rho}|[\boldsymbol{c}]_t)P_t(\boldsymbol{n}_i+\mathbf S_{i,\rho})}
        \,,
        \label{eq:EPRgeneral}
    \end{equation}
    given in units of Boltzmann constant $k_{\mathrm B}$.
    
%%%%%%%%%%%%%%%%%%%%%%%%%%%%%%%%%%%%%%%%%%%%%%%%%%%%%%%%%%%%
%%%%%%%%%%%%%%%%%%%%%%%%%%%%%%%%%%%%%%%%%%%%%%%%%%%%%%%%%%%%
%%%%%%%%%%%%%%%%%%%%%%%%%%%%%%%%%%%%%%%%%%%%%%%%%%%%%%%%%%%%

% \newpage
    \section{Hybrid Chemical Reaction Networks}
	\label{sec:hybridDynamics}
    
%%%%%%%%%%%%%%%%%%%%%%%%%%%%%%%%%%%%%%%%%%%%%%%%%%%%%%%%%%%%
	\subsection{Setup}
    \label{sec:hCRNsetup}

    Hybrid CRNs are defined here as closed CRNs
    where the species~$\{i\}$ can be classified,
    based on their abundances,
    into
    \textit{low-abundant} species (labeled~$x$)
    and 
    \textit{high-abundant} species (labeled~$y$).
    % -
    In a large volume $\Omega \to \infty$,
    the low-abundant species have finite molecule numbers $n_x$,
    while the high-abundant species have finite concentrations,
    defined as $[y] \equiv n_y / \Omega$.
    % -

    % -
    In this framework, the chemical equation~\eqref{eq:CRNopen} specializes to
    \begin{equation}\small
        \sum_x \nu_{x,\rho} Z_x + \sum_y \nu_{y,\rho} Z_y 
        \ch{<=>[$\rho$][$-\rho$]}  
        \sum_x \nu_{x,-\rho} Z_x +\sum_y \nu_{y,-\rho} Z_y 
        \,,
        \label{eq:CRNinternal}
    \end{equation}
    and the set of reactions $\mathcal R$ can be split into 
    \textit{continuous} $\mathcal R_c$ and \textit{discrete} $\mathcal R_d$ reactions. 
    % -
    Each reaction $\rho$ is a continuous reaction 
    if it does not affect the abundance of the low-abundant species.~\cite{menz12}
    Namely,
    \begin{equation}
        \mathcal R_c \equiv \{\rho \in \mathcal R \,|\, \mathbf S_{x,\rho}=\mathbf 0\}\,,
        \label{eq:DefContinuousReactions}
    \end{equation}
    while $\mathbf S_{y,\rho}\neq \mathbf0$ $\forall \rho \in \mathcal R_c$.
    % -
    In contrast, the discrete reactions
    \begin{equation}
        \mathcal R_d \equiv \mathcal R \setminus \mathcal R_c.
        \label{eq:DefDiscreteReactions}
    \end{equation}
    affect the abundance of the low-abundant species.~\cite{menz12} 
    Namely,
    $\mathbf S_{x,\rho} \neq \mathbf 0$ $\forall \rho \in \mathcal R_d$.
    % -
    The terms continuous and discrete reactions are chosen 
    anticipating the dynamics that they will lead to for the 
    high-abundant and low-abundant species, respectively, 
    in hybrid CRNs (see Subs.~\ref{sec:HybridGeneral}).

%%%%%%%%%%%%%%%%%%%%%%%%%%%%%%%%%%%%%%%%%%%%%%%%%%%%%%%%%%%%

    \subsection{Dynamics}
    \label{sec:HybridGeneral}

    The hybrid dynamics is derived
    (as summarized in Appendix~\ref{appendix:hybrid})
    by using a systematic {\col(single time-scale)} perturbation analysis
    in the limit $\Omega \to \infty$ 
    of the CME~\eqref{eq:CME}
    with finite molecule numbers $\boldsymbol{n}_x$ 
    and 
    finite concentrations $[\boldsymbol{y}]\equiv(\dots, [y] ,\dots)$,
    called \textit{partial macroscopic limit}.~\cite{menz12, wink17}
    % - 
    The procedure leads to a closed dynamical description only under the assumption that the (scaled) reaction rates 
    \begin{equation}
    r_\rho(\boldsymbol n_x,[\boldsymbol{y}]) \equiv \begin{cases}
        R_\rho(\boldsymbol n_x,\Omega[\boldsymbol{y}])/\Omega\,,& \rho \in \mathcal R_c\\
        R_\rho(\boldsymbol n_x,\Omega[\boldsymbol{y}])\,,& \rho \in \mathcal R_d
    \end{cases}
    \label{eq:def_rates_hCRNs}
    \end{equation}
    with $R_\rho(\boldsymbol n_x,\boldsymbol n_y) \equiv  R_\rho(\boldsymbol n_i)|_{\boldsymbol n_i = (\boldsymbol n_x, \boldsymbol n_y)}$
    and $\boldsymbol n_y =\Omega[\boldsymbol{y}]$,
    satisfy 
    \begin{equation}
        r_\rho(\boldsymbol n_x,[\boldsymbol{y}]) = \mathcal O(1)\text{  for  }\Omega \to \infty\,.
        \label{eq:ReactionRatesScaling}
    \end{equation}
    Note that the physical meaning of the scaling in Eq.~\eqref{eq:ReactionRatesScaling}
    will become clear at the end of this section.
    % -
    % -

    % -
    The resulting dynamics for the concentrations of high-abundant species is, 
    to leading order, 
    a piecewise deterministic process 
    described by the following reaction rate equation~(RRE),
    \begin{equation}
%    {\col
 %       \begin{cases}
            \partial_t [\boldsymbol{y}|\boldsymbol{n}_x]_t = 
            \sum\limits_{\rho \in \mathcal R_c} 
            \mathbf S_{y,\rho} \, 
            r_\rho(\boldsymbol{n}_x,[\boldsymbol{y}|\boldsymbol{n}_x]_t)\\
  %          \phantom{\partial_t} [\boldsymbol y|\boldsymbol n_x]_0 = [{\boldsymbol y}]_0
   %     \end{cases}
            \,,
    %}
        \label{eq:RRElot}
    \end{equation}
    where $[\boldsymbol{y}|\boldsymbol{n}_x]_t$ 
    is the vector of the concentrations of the high-abundant species at time $t$, 
    given the molecule numbers $\boldsymbol{n}_x$ of the low-abundant species, {\col evolving from a given initial condition $[\boldsymbol{y}|\boldsymbol{n}_x]_{t=0} = [\boldsymbol y]_0$}. 
    % -
    Physically, 
    the RRE~\eqref{eq:RRElot} determines 
    a \textit{continuous} evolution of the concentrations of the high-abundant species 
    depending on the low-abundant species.
    % -
    For this reason, the $\mathcal R_c$ reactions are called continuous reactions.
    % -
    % -
    On the other hand, 
    the resulting dynamics for the molecule numbers of the low-abundant species is, 
    to leading order,
    a jump process 
    described by the following~CME
    \begin{equation}\small
        \begin{split}
            \partial_t p_t(\boldsymbol{n}_x) = 
            \sum\limits_{\rho \in \mathcal R_d} 
            \big\{
                &r_\rho(\boldsymbol{n}_x-\mathbf S_{x,\rho},[\boldsymbol{y}|\boldsymbol{n}_x-\mathbf S_{x,\rho}]_t)p_t(\boldsymbol{n}_x-\mathbf S_{x,\rho}) \\ 
                & - r_\rho(\boldsymbol{n}_x,[\boldsymbol{y}|\boldsymbol{n}_x]_t)p_t(\boldsymbol{n}_x) \}\,{\col ,}
        \end{split}
        \label{eq:CMElot}
    \end{equation}
    {\col where $p_t(\boldsymbol n_x)$ is the probability of there being $\boldsymbol n_x$ molecules of the low-abundant species at time $t$,
    evolving from a given initial condition $p_{t=0}(\boldsymbol n_x)$.}
    % -
    Physically,
    the CME~\eqref{eq:CMElot} determines
    a \textit{discrete} variations of the molecule numbers of the low-abundant species
    depending on the high-abundant species.
    % -
    For this reason, the $\mathcal R_d$ reactions are called discrete reactions.
    % -
    % -
    Note that 
    $p_t(\boldsymbol{n}_x)$ is the leading order approximation of 
    the marginalized joint probability distribution $P_t(\boldsymbol{n}_x,\boldsymbol{n}_y)$, 
    i.e., $p_t(\boldsymbol{n}_x) \simeq\sum_{\boldsymbol{n}_y}P_t(\boldsymbol{n}_x,\boldsymbol{n}_y)$ 
    for $\Omega \to \infty$ (see Appendix \ref{appendix:hybrid}).
    % -

    % -
    In summary, 
    the molecule numbers $\boldsymbol{n}_x$ evolve 
    according to a jump process (described by the CME~\eqref{eq:CMElot})
    whose reaction rates depend on the concentrations of the high-abundant species.
    % - 
    During the dwell time between one jump and the next one,
    the concentrations $[\boldsymbol y]$ of the high-abundant species evolve 
    according to a deterministic process (described by RRE~\eqref{eq:RRElot})
    whose rates depend on the molecule number of the low-abundant species. 
    % -
    % -
    Furthermore,
    since the reaction rates featuring
    the CME~\eqref{eq:CMElot} and the RRE~\eqref{eq:RRElot} are of the same order
    because of the scaling condition in Eq.~\eqref{eq:ReactionRatesScaling},
    the molecule numbers $\boldsymbol{n}_x$ 
    and the concentrations  $[\boldsymbol y]$ 
    evolve on the same time scale. 
    {\col 
    For this reason, the partial macroscopic limit corresponds to 
    a single time-scale perturbation analysis 
    that cannot resolve reactions 
    occurring on multiple time-scales whose rates do not satisfy 
    the scaling condition in Eq.~\eqref{eq:ReactionRatesScaling}.
    }
    
    % - - - -
    
    % - - - -

    % - - - -
    
%%%%%%%%%%%%%%%%%%%%%%%%%%%%%%%%%%%%%%%%%%%%%%%%%%%%%%%%%%%%
%%%%%%%%%%%%%%%%%%%%%%%%%%%%%%%%%%%%%%%%%%%%%%%%%%%%%%%%%%%%

\section{Hybrid dynamics of elementary reactions} \label{sec:HDele}

%    \subsection{Dynamics: Mass-action kinetics}
    \subsection{Mass-action kinetics}
    \label{sec:MassAction}
    The hybrid dynamics 
    summarized in Subs.~\ref{sec:HybridGeneral} and App.~\ref{appendix:hybrid}
    is valid for any reaction rates $R_\rho(\boldsymbol{n}_x,\boldsymbol{n}_y)$ 
    satisfying  the scaling condition in Eq.~\eqref{eq:ReactionRatesScaling}. 
    % -
    Here, as well as throughout the paper, 
    we focus on elementary reactions in ideal dilute mixtures
    satisfying mass-action kinetics,
    whose reaction rates~\eqref{eq:MassActionOpen} can be rewritten as
    \begin{equation}\small
        R_\rho(\boldsymbol{n}_x,\boldsymbol{n}_y) 
        = 
        \Omega k_\rho 
        \prod_x \frac{n_x!}{(n_x-\nu_{x,\rho})!\Omega^{\nu_{x,\rho}}}
        \prod_y \frac{n_y!}{(n_y-\nu_{y,\rho})!\Omega^{\nu_{y,\rho}}}\,,
        \label{eq:MassActionSplitted}
    \end{equation}
    when considering closed CRNs and
    the splitting of the chemical species into
    low- and high-abundant ones.
    % -
    % - 
    Since the volume dependence of the reaction rates 
    in Eq.~\eqref{eq:MassActionSplitted}
    is determined by the stoichiometric coefficients 
    $\{\nu_{x,\rho}\}$ and $\{\nu_{y,\rho}\}$,
    % -
    only specific stoichiometries ensure that
    the scaling condition in Eq.~\eqref{eq:ReactionRatesScaling} is satisfied.
    % -
    In particular, 
    by comparing the $\Omega$-dependence in Eq.~\eqref{eq:MassActionSplitted}
    with Eqs.~\eqref{eq:def_rates_hCRNs} and~\eqref{eq:ReactionRatesScaling},
    we conclude that
    all discrete reactions must be unimolecular reactions 
    in the low-abundant species, i.e.,  
    \begin{equation}
        Z_x + \sum_y \nu_{y,\rho} Z_y 
        \ch{<=>[$\rho$][$-\rho$]} 
        Z_{x^\prime} + \sum_y \nu_{y,-\rho} Z_y
        \,,
        \label{eq:ConstraintDiscrete}
    \end{equation}
    for all $\rho\in \mathcal R_d$ 
    and with $x \neq x'$ by definition of discrete reaction.
    % -
    Furthermore, 
    all continuous reactions must not involve the low-abundant species, i.e.,
    \begin{equation}
        \sum_y \nu_{y,\rho} Z_y 
        \ch{<=>[$\rho$][$-\rho$]}
        \sum_y \nu_{y,-\rho} Z_y\,,
        \label{eq:ConstraintContinuous}
    \end{equation}
    for all $\rho\in \mathcal R_c$.
    % -
    % -
    This crucially implies that only hybrid CRNs 
    with discrete and continuous reactions satisfying
    the stoichiometric conditions 
    in Eqs.~\eqref{eq:ConstraintDiscrete} and~\eqref{eq:ConstraintContinuous} 
    {\col(and illustrated in Fig.~\ref{fig:3})}
    have a well-defined dynamical description
    (given in Sec.~\ref{sec:HybridGeneral})
    in the partial macroscopic limit.
    % -

    % -
    \textit{Remark.}
    If there are no continuous reactions, i.e., $\mathcal R_c = \emptyset$,
    the hybrid dynamics can also be derived when  
    \begin{equation}
        r_\rho(\boldsymbol{n}_x, [\boldsymbol{y}]) = 
        R_\rho(\boldsymbol{n}_x,\Omega [\boldsymbol{y}]) / \Omega^{\alpha}
        \,,
    \end{equation}
    where $\alpha$ is an integer number equal for all $\rho \in \mathcal R_d$.
    Indeed, the scaling condition in Eq.~\eqref{eq:ReactionRatesScaling} is restored
    by rescaling time, i.e., by defining $\tau \equiv \Omega^{\alpha} t$.
    % -
    
    % -
    {\col
    \textit{Remark.}
    When the stoichiometric conditions
    in Eqs.~\eqref{eq:ConstraintDiscrete} and~\eqref{eq:ConstraintContinuous}
    are not satisfied, 
    reaction occur on multiple time-scales
    (proportional to powers of $1/\Omega$)
    and, therefore,
    cannot be resolved with a single time-scale perturbation analysis such as
    the partial macroscopic limit.
    }

%%%%%%%%%%%%%%%%%%%%%%%%%%%%%%%%%%%%%%%%%%%%%%%%%%%%%%%%%%%%
     
	\subsection{Emergence of open CRNs 
    with autonomous chemostats}
    \label{sec:HybridChemostat}

    In this section,
    we prove that the dynamical description of open CRNs with autonomous chemostats
    emerges from the hybrid dynamics with no continuous reactions.
    %-
    Indeed, in the absence of continuous reactions, 
    the RRE~\eqref{eq:RRElot} for the high-abundant species boils down to
    \begin{equation}
        \partial_t [\boldsymbol{y}]_t = 0
        \,,
    \end{equation}
    leading to constant concentrations, i.e.,
    $[\boldsymbol{y}]_t = [\boldsymbol{y}]_0$ $\forall t \geq0$, 
    with $[\boldsymbol{y}]_0$ being the vector 
    of the initial concentrations of the high-abundant species.
    % -
    By plugging $[\boldsymbol{y}]_0$ into Eq.~\eqref{eq:CMElot},
    the CME for the low-abundant species becomes
    \begin{equation}
        \begin{split}
            \partial_t p_t(\boldsymbol{n}_x) = 
            \sum\limits_{\rho \in \mathcal R_d} 
            \big\{
            & r_\rho(\boldsymbol{n}_x-\mathbf S_{x,\rho},[\boldsymbol y]_0)
            p_t(\boldsymbol{n}_x-\mathbf S_{x,\rho}) 
            \\ &- r_\rho(\boldsymbol{n}_x,[\boldsymbol y]_0)p_t(\boldsymbol{n}_x) 
            \big\}
            \,,
        \end{split}\label{eq:CMEchemostat}
    \end{equation}
    where the reaction rates are given, to leading order, by
    \begin{equation}
        r_\rho(\boldsymbol{n}_x, [\boldsymbol{y}]_0) \simeq
        \underbrace{
        \Omega k_\rho 
        \prod_x \frac{n_x!}{(n_x-\nu_{x,\rho})!\Omega^{\nu_{x,\rho}}}
        \prod_y [y]_0^{\nu_{y,\rho}} 
        }_{
        = R_\rho(\boldsymbol{n}_x|[\boldsymbol{y}]_0)}\,,
        \label{eq:MassActionLOT}
    \end{equation}
    with $\nu_{x,\rho} \neq 0 $ only for one specific species 
    according to the stoichiometric conditions in Eq.~\eqref{eq:ConstraintDiscrete}.
    % -
    By now mapping 
    the molecule number $\boldsymbol n_x$ and 
    the concentrations $[\boldsymbol y]_0$ 
    into
    the molecule number of the internal species $\boldsymbol n_i$
    and the concentrations of chemostats $[\boldsymbol c]$
    of open CRNs, respectively,
    the CME~\eqref{eq:CMEchemostat} of hybrid CRNs
    becomes equivalent to
    the CME~\eqref{eq:CME} of open CRNs
    with autonomous chemostats.
    % -
    This physically means that, to leading order,
    the hybrid dynamics describes the dynamics of open CRNs
    with the low- and high-abundant species playing the role 
    of the internal species and autonomous chemostats, respectively.
    % -

%%%%%%%%%%%%%%%%%%%%%%%%%%%%%%%%%%%%%%%%%%%%%%%%%%%%%%%%%%%%

	\subsection{Emergence of open CRNs 
    with evolving chemostats}
    \label{sec:HybridEvolve}
    In this section,
    we prove that the dynamical description of open CRNs 
    with evolving chemostats
    emerges from the hybrid dynamics with continuous reactions.
    % -
    Indeed, in the presence of continuous reactions
    satisfying the stoichiometric conditions in Eq.~\eqref{eq:ConstraintContinuous},
    the RRE~\eqref{eq:RRElot} for the high-abundant species becomes
    \begin{equation}
        \partial_t [\boldsymbol{y}]_t 
        = \sum\limits_{\rho \in \mathcal R_c} 
        \mathbf S_{y,\rho}\, r_\rho([\boldsymbol{y}]_t)
        \,,
        \label{eq:CMassActionevolve}    
    \end{equation}
    with the reaction rates given, to leading order, by
    \begin{equation}
        r_\rho([\boldsymbol{y}]_t) \simeq 
        k_\rho \prod_y [{y}]_t^{\nu_{y,\rho}}
        = R_\rho([\boldsymbol{y}]_t) / \Omega
        \,.
        \label{eq:RREevolve}
    \end{equation}
    This implies that 
    the evolution of the concentrations $[\boldsymbol{y}]_t$ 
    is independent of the low-abundant species
    (unlike the general RRE~\eqref{eq:RRElot} which does not account 
    for the constraints resulting from mass-action kinetics).
    % -
    By plugging $[\boldsymbol{y}]_t$ into Eq.~\eqref{eq:CMElot},
    the CME for the low-abundant species becomes
    \begin{equation}
        \begin{split}
           \partial_t p_t(\boldsymbol{n}_x) = 
            \sum\limits_{\rho \in \mathcal R_d} 
            \big\{
            & r_\rho(\boldsymbol{n}_x-\mathbf S_{x,\rho},[\boldsymbol y]_t)
            p_t(\boldsymbol{n}_x-\mathbf S_{x,\rho}) 
            \\ &- r_\rho(\boldsymbol{n}_x,[\boldsymbol y]_t)p_t(\boldsymbol{n}_x) 
            \big\}
            \,,
        \end{split}\label{eq:CMEevolve}
    \end{equation}
    where the reaction rates are given, to leading order, by
    \begin{equation}
        r_\rho(\boldsymbol{n}_x, [\boldsymbol{y}]_t) \simeq
        \underbrace{
        \Omega k_\rho 
        \prod_x \frac{n_x!}{(n_x-\nu_{x,\rho})!\Omega^{\nu_{x,\rho}}}
        \prod_y [y]_t^{\nu_{y,\rho}} 
        }_{
        = R_\rho(\boldsymbol{n}_x|[\boldsymbol{y}]_t)}\,,
        \label{eq:MassActionEvolve}
    \end{equation}
    with $\nu_{x,\rho} \neq 0 $ only for one specific species 
    according to the stoichiometric conditions in Eq.~\eqref{eq:ConstraintDiscrete}.
    % -
    By now mapping again
    the molecule number $\boldsymbol n_x$ and 
    the concentrations $[\boldsymbol y]_t$ 
    into
    the molecule number of the internal species $\boldsymbol n_i$
    and the concentrations of chemostats $[\boldsymbol c]_t$
    of open CRNs, respectively,
    the CME~\eqref{eq:CMEevolve} of hybrid CRNs
    %for the low abundant species in hybrid CRNs 
    becomes equivalent to
    the CME~\eqref{eq:CME} of open CRNs
    with evolving chemostats.
    % -
    This physically means that, to leading order,
    the hybrid dynamics describes the dynamics of open CRNs
    with the low- and high-abundant species playing the role 
    of the internal species and evolving chemostats, respectively.
    Furthermore,
    the continuous reactions act as the external processes controlling 
    the concentrations of the chemostats.
    % -
    
%%%%%%%%%%%%%%%%%%%%%%%%%%%%%%%%%%%%%%%%%%%%%%%%%%%%%%%%%%%%
%%%%%%%%%%%%%%%%%%%%%%%%%%%%%%%%%%%%%%%%%%%%%%%%%%%%%%%%%%%%

    \section{Hybrid Thermodynamics of Elementary Reactions}
	\label{sec:TD}

	\subsection{Local detailed balance condition}\label{sec:HDldb}

    We prove here that the dynamics of hybrid CRNs
    (Subs.~\ref{sec:HybridGeneral}) 
    is 
    thermodynamically consistent, 
    namely, 
    it satisfies a local detailed balance condition
    in the partial macroscopic limit.
    % -
    % -
    To do so, we rewrite the general
    local detailed balance condition in Eq.~\eqref{eq:LDBopen}
    for closed CRNs
    where the chemical species are split into
    low- and high-abundant species,
    \begin{equation}
        \ln\frac{
        R_{\rho}(\boldsymbol{n}_x,\boldsymbol{n}_y)
        }{
        R_{-\rho}(\boldsymbol{n}_x+\mathbf S_{x,\rho},\boldsymbol{n}_y +\mathbf S_{y,\rho})}
        = 
        -  
        \big\{ 
        \Delta_\rho g(\boldsymbol n_x) 
        + 
        \Delta_\rho g(\boldsymbol n_y) 
        \big\}\,.
        \label{eq:LDBinternal}
    \end{equation}
    % -

    % -
    We now take the partial macroscopic limit of Eq.~\eqref{eq:LDBinternal}.
    % - 
    % -
    By using Eq.~\eqref{eq:def_rates_hCRNs} 
    together with 
    Eq.~\eqref{eq:MassActionLOT} or Eq.~\eqref{eq:MassActionEvolve}
    (depending on whether continuous reactions are present or not)
    on the left-hand side of Eq.~\eqref{eq:LDBinternal},
    we obtain
        \begin{equation}
         \ln\frac{
         R_{\rho}(\boldsymbol{n}_x,\boldsymbol{n}_y)
         }{
         R_{-\rho}(\boldsymbol{n}_x+\mathbf S_{x,\rho},\boldsymbol{n}_y +\mathbf S_{y,\rho})}
         \simeq 
         \ln\frac{R_{\rho}(\boldsymbol{n}_x|[\boldsymbol{y}])
         }{
         R_{-\rho}(\boldsymbol{n}_x+\mathbf S_{x,\rho}|[\boldsymbol{y}])}
         \,,
         \label{eq:LDBlhs}
    \end{equation}
    to leading order.
    Equation~\eqref{eq:LDBlhs} boils down to 
    \begin{equation}
         \ln\frac{
         R_{\rho}(\boldsymbol{n}_y)
         }{
         R_{-\rho}(\boldsymbol{n}_y +\mathbf S_{y,\rho})}
         \simeq 
         \ln\frac{R_{\rho}([\boldsymbol{y}])
         }{
         R_{-\rho}([\boldsymbol{y}])}
         \,,
        \label{eq:LDBlhscontinous}
    \end{equation}
    when considering continuous reactions $\rho \in \mathcal R_c$,
    where $\nu_{x,\rho} = \nu_{x,-\rho} = 0$ $\forall x$
    because of the stoichiometric conditions in Eq.~\eqref{eq:ConstraintContinuous}.
    % - 

    % -
    We now take the partial macroscopic limit 
    of the right-hand side of Eq.~\eqref{eq:LDBinternal}.
    This corresponds to
    taking the macroscopic limit
    of the variation of the Gibbs free energy along reaction $\rho$ 
    due to each high-abundant species, i.e., $\Delta_\rho g(n_y)$.
    We thus obtain
    \begin{equation}\small
        \begin{split}
            \Delta_\rho g(n_y) 
            & =
            (\mu^\circ_y - \ln n_s) S_{y,\rho} 
            +\ln (n_y + S_{y,\rho})! - \ln n_y!\\
            & \approx 
            (\mu^\circ_y - \ln n_s) S_{y,\rho} + 
            {\col \Omega \{u \ln (\Omega u) - u\}\Big|_{[y]}^{[y] + S_{y,\rho}/\Omega }}\\
            & \approx 
            (\mu^\circ_y - \ln n_s) S_{y,\rho} + 
            \ln(\Omega [y]) S_{y,\rho}\\
            & =
            \big(\mu^\circ_y + \ln ([y]/[s])\big) S_{y,\rho} 
            \equiv 
            \mu([y]) \, S_{y,\rho}\,,
        \end{split}
        \label{eq:LDBrhs}
    \end{equation}
    by
    using Stirling's formula, i.e., $\ln n_y! \approx n_y \ln n_y - n_y$, {\col for the first approximation,}
    a Taylor expansion for {\col ${S_{y,\rho} /\Omega \ll 1}$ for the second approximation},
    and by identifying the chemical potential of the high-abundant species as
    $\mu([y]) \equiv \mu^\circ_y + \ln ([y]/[s])$ {\col in the final step}.
    % -
    
    % -
    Finally, by plugging Eqs.~\eqref{eq:LDBlhs} and~\eqref{eq:LDBrhs} 
    into Eq.~\eqref{eq:LDBinternal}, 
    we arrive at the following local detailed balance condition
    \begin{equation}
        \ln 
        \frac{
        R_{\rho}(\boldsymbol{n}_x|[\boldsymbol{y}])
        }{
        R_{-\rho}(\boldsymbol{n}_x+\mathbf S_{x,\rho}|[\boldsymbol{y}])
        } 
        = 
        -
        [
        \Delta_\rho g(\boldsymbol{n}_x) 
        + \boldsymbol{\mu}([\boldsymbol{y}])\cdot \mathbf S_{y,\rho}]
        \,,
    \end{equation}
    which boils down to 
    \begin{equation}
        \ln 
        \frac{
        R_{\rho}([\boldsymbol{y}])
        }{
        R_{-\rho}([\boldsymbol{y}])
        } 
        = 
        -\boldsymbol{\mu}([\boldsymbol{y}])\cdot \mathbf S_{y,\rho}
        \,,
    \end{equation}
    for continuous reactions $\rho \in \mathcal R_c$ {\col due to the stoichiometric conditions in Eq.~\eqref{eq:ConstraintContinuous}}.
    % -
    % -
    In practice, 
    the local detailed balance condition~\eqref{eq:LDBinternal} of hybrid CRNs
    converges, in the partial macroscopic limit,
    to the local detailed balance condition~\eqref{eq:LDBopen} of open CRNs
    with the low- and high-abundant species playing the role 
    of internal species and chemostats, respectively. 
    % -
    
	\subsection{Entropy production rate}\label{sec:HDepr}

    We derive here the entropy production rate for hybrid CRNs.
    To do so, we take the partial macroscopic limit of
    the entropy production rate in Eq.~\eqref{eq:EPRgeneral},
    which can be rewritten as
    \begin{equation}\small
    \begin{split}
        \braket{\dot \Sigma}_t = &
        \sum\limits_{\rho,\boldsymbol{n}_x,\boldsymbol{n}_y} 
        R_{\rho}(\boldsymbol{n}_x, \boldsymbol{n}_y)P_t(\boldsymbol{n}_x, \boldsymbol{n}_y)\\
        & \times 
        \ln \frac{
        R_{\rho}(\boldsymbol{n}_x, \boldsymbol{n}_y)P_t(\boldsymbol{n}_x, \boldsymbol{n}_y)
        }{
        R_{-\rho}(\boldsymbol{n}_x+\mathbf S_{x,\rho}, \boldsymbol{n}_y+\mathbf S_{y,\rho})P_t(\boldsymbol{n}_x+\mathbf S_{x,\rho}, \boldsymbol{n}_y+\mathbf S_{y,\rho})}
        \,,
    \end{split}
    \label{eq:EPRnxny}
    \end{equation}
    when considering closed CRNs and
    the splitting of the chemical species into
    low- and high-abundant ones.
    % -
    Furthermore, like in Sec.~\ref{sec:hybridDynamics},
    we consider separately hybrid CRNs with only discrete reactions
    and hybrid CRNs with also continuous reactions.

    \subsubsection{Hybrid CRNs without continuous reactions}\label{subsub:EPRdiscrete}

    From a dynamical standpoint, 
    in hybrid CRNs without continuous reactions
    the concentrations $[\boldsymbol y]$ of the high-abundant species are constant in time,
    i.e., $[\boldsymbol{y}]_t = [\boldsymbol{y}]_0$ $\forall t\geq0$,
    and play the role of autonomous chemostats
    (see Subs.~\ref{sec:HybridChemostat}).
    % -
    % -
    From a thermodynamic standpoint,
    only the discrete reactions dissipate and, 
    by taking the partial macroscopic limit of Eq.~\eqref{eq:EPRnxny}
    as formally done in App.~\ref{appendix:epr},
    the entropy production rate converges, to leading order, to
    \begin{equation}
    \begin{split}
        \braket{\dot \Sigma}_t 
        \simeq 
         &
         \sum_{\rho\in \mathcal R_d,\boldsymbol{n}_x}
         R_{\rho}(\boldsymbol{n}_x|[\boldsymbol{y}]_0)p_t(\boldsymbol{n}_x)\\
        &
         \times
         \ln \frac{
         R_{\rho}(\boldsymbol{n}_x|[\boldsymbol{y}]_0)p_t(\boldsymbol{n}_x)
         }{
         R_{-\rho}(\boldsymbol{n}_x+\mathbf S_{x,\rho}|[\boldsymbol{y}]_0)p_t(\boldsymbol{n}_x+\mathbf S_{x,\rho})}
         \,.
    \end{split}
    \label{eq:EPRChemostat}
    \end{equation}
    % -
    % - 
    The entropy production rate~\eqref{eq:EPRChemostat}
    of hybrid CRNs without continuous reactions 
    is equivalent to the entropy production rate~\eqref{eq:EPRgeneral}
    of open CRNs with autonomous chemostats
    with
    the low- and high-abundant species playing the role of 
    the internal species and autonomous chemostats, respectively.
    % -
    Hence, the mapping between hybrid CRNs and open CRNs holds at the 
    dynamical level as well as at the thermodynamic one.

    \subsubsection{Hybrid CRNs with continuous reactions}\label{sec:EPRcontinuous}
    From a dynamical standpoint, 
    in hybrid CRNs with continuous reactions
    the concentrations $[\boldsymbol y]_t$ of the high-abundant species evolve in time according to the RRE~\eqref{eq:CMassActionevolve}
    and play the role of evolving chemostats
    (see Subs.~\ref{sec:HybridEvolve}).
    % -
    % -
    From a thermodynamic standpoint,
    the entropy production rate can be written as the sum of two contributions
    according to
    \begin{equation}
        \braket{\dot \Sigma}_t 
        = 
        \braket{\dot \Sigma_d}_t + 
        \braket{\dot \Sigma_c}_t 
        \,.
    \end{equation}
    where $\braket{\dot \Sigma_d}_t$ and $\braket{\dot \Sigma_c}_t$
    are the entropy production rate of the 
    discrete and continuous reactions, respectively.
    % -
    By now taking the partial macroscopic limit of 
    $\braket{\dot \Sigma_d}_t$ and $\braket{\dot \Sigma_c}_t$
    as formally done in App.~\ref{appendix:epr},
    we obtain, to leading order,
    \begin{equation}
    \begin{split}
        \braket{\dot \Sigma_d}_t 
        \simeq 
         &
         \sum_{\rho\in \mathcal R_d,\boldsymbol{n}_x}
         R_{\rho}(\boldsymbol{n}_x|[\boldsymbol{y}]_t)p_t(\boldsymbol{n}_x)\\
        &
         \times
         \ln \frac{
         R_{\rho}(\boldsymbol{n}_x|[\boldsymbol{y}]_t)p_t(\boldsymbol{n}_x)
         }{
         R_{-\rho}(\boldsymbol{n}_x+\mathbf S_{x,\rho}|[\boldsymbol{y}]_t)p_t(\boldsymbol{n}_x+\mathbf S_{x,\rho})}
         \,,
    \end{split}
    \label{eq:EPRevolveD}
    \end{equation}
    for the discrete reactions and
    \begin{equation}
        \braket{\dot \Sigma_c}_t 
        \simeq
        \sum\limits_{\rho\in \mathcal R_c} 
        R_{\rho}([\boldsymbol{y}]_t)
        \ln 
        \frac{R_{\rho}([\boldsymbol{y}]_t)}{R_{-\rho}([\boldsymbol{y}]_t)}
        \,,
        \label{eq:EPRevolveC}
    \end{equation}
    for the continuous reactions.
    % -
    % -
    
    Let us now examine the meaning of the two expressions 
    in Eqs.~\eqref{eq:EPRevolveD} and~\eqref{eq:EPRevolveC}.
    % -
    On the one hand, as in Subs.~\ref{subsub:EPRdiscrete},
    the entropy production rate~\eqref{eq:EPRevolveD}
    is equivalent to the entropy production rate~\eqref{eq:EPRgeneral}
    of open CRNs with evolving chemostats
    and 
    the mapping between hybrid CRNs and open CRNs holds at the 
    dynamical level as well as at the thermodynamic one.
    On the other hand,
    the entropy production rate~\eqref{eq:EPRevolveC} 
    accounts for the dissipation of the reactions controlling the evolution 
    of the continuous species.
    % -
    Since these reactions play the role of the external processes controlling 
    the chemostats when mapping hybrid CRNs into open CRNs,
    the entropy production $\braket{\dot \Sigma_c}_t $ thus quantifies the dissipation of
    the external processes
    (which is never accounted for in standard modeling of open CRNs).
    % -
    
    % -
    We furthermore stress that $\braket{\dot \Sigma_d}_t$ and $\braket{\dot \Sigma_c}_t$
    are of different order in $\Omega$.
    % -
    The former is $\Omega$-independent
    since, according to the stoichiometric conditions in
    Eq.~\eqref{eq:ConstraintDiscrete},
    the discrete reactions are unimolecular in the low-abundant species 
    and, therefore, $R_{\rho}(\boldsymbol{n}_x|[\boldsymbol{y}]_t) = \mathcal O(1)$ 
    (see mass-action kinetics in Eq.~\eqref{eq:MassActionOpen}).
    % -
    The latter scales linearly in $\Omega$
    since, according to the stoichiometric conditions in 
    Eq.~\eqref{eq:ConstraintContinuous},
    the continuous reactions do not involve low-abundant species
    and, therefore, $R_{\rho}([\boldsymbol{y}]_t) = \mathcal O(\Omega)$ 
    (see mass-action kinetics in Eq.~\eqref{eq:MassActionOpen}).
    % -
    Hence, as long as both discrete and continuous reactions do not equilibrate,
    the entropy production rate of the continuous reactions 
    dominates 
    the entropy production rate of the discrete reactions.

%%%%%%%%%%%%%%%%%%%%%%%%%%%%%%%%%%%%%%%%%%%%%%%%%%%%%%%%%%%%
%%%%%%%%%%%%%%%%%%%%%%%%%%%%%%%%%%%%%%%%%%%%%%%%%%%%%%%%%%%%

    % \newpage\text{}
    % \newpage
	\section{Illustrative examples}
	\label{sec:examples}
    
    We now compare the exact dynamics and thermodynamics of closed CRNs
    with the corresponding hybrid descriptions
    in the partial macroscopic limit
    for two prototypical examples.
	
	\subsection{Michaelis-Menten}\label{sec:MM}

    \begin{figure*}
	\centering
	\subfloat[\label{sfig:1a}]{%
		\includegraphics[width=0.32\textwidth]{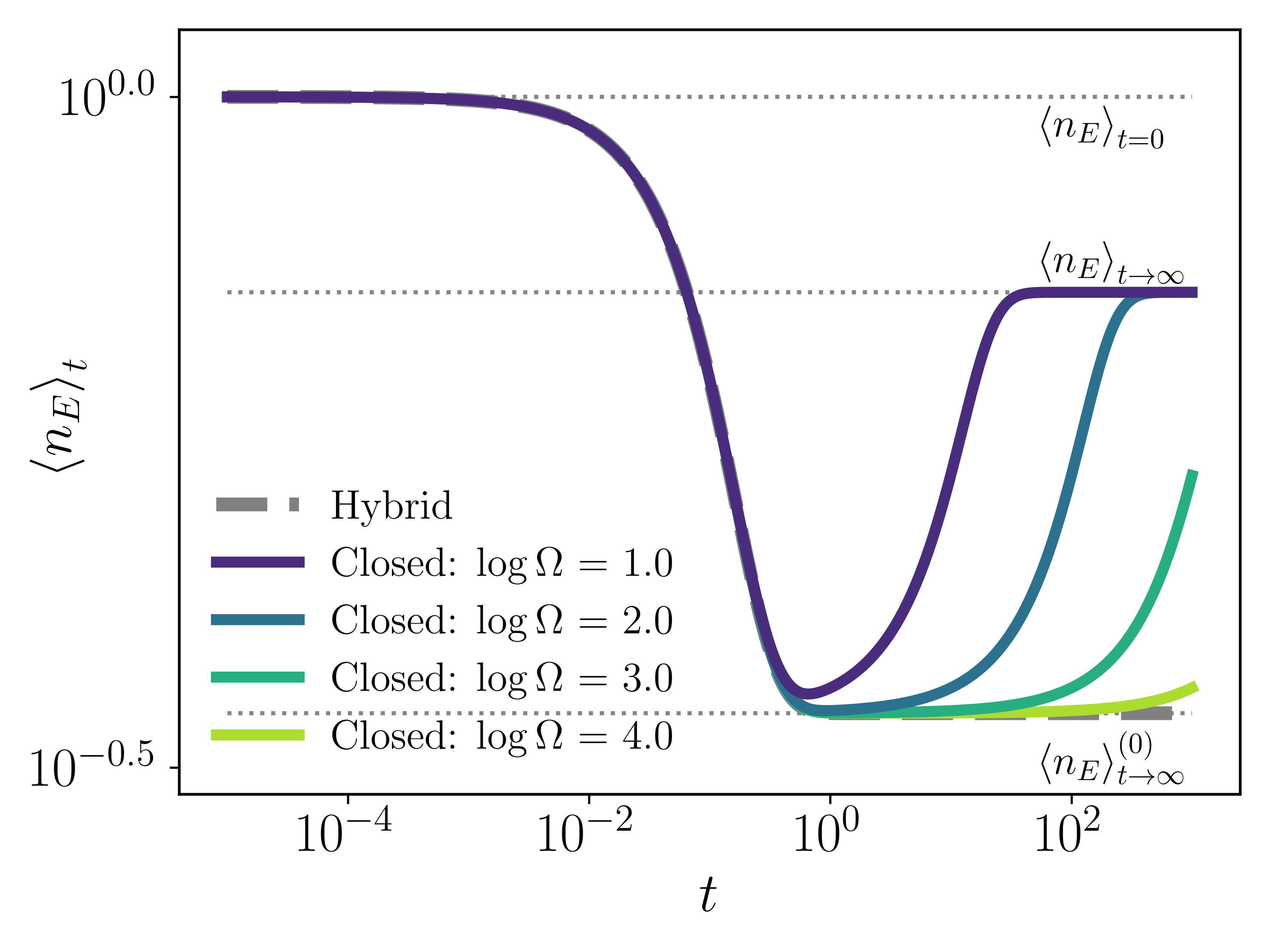}}
	\subfloat[\label{sfig:1b}]{%
		\includegraphics[width=0.32\textwidth]{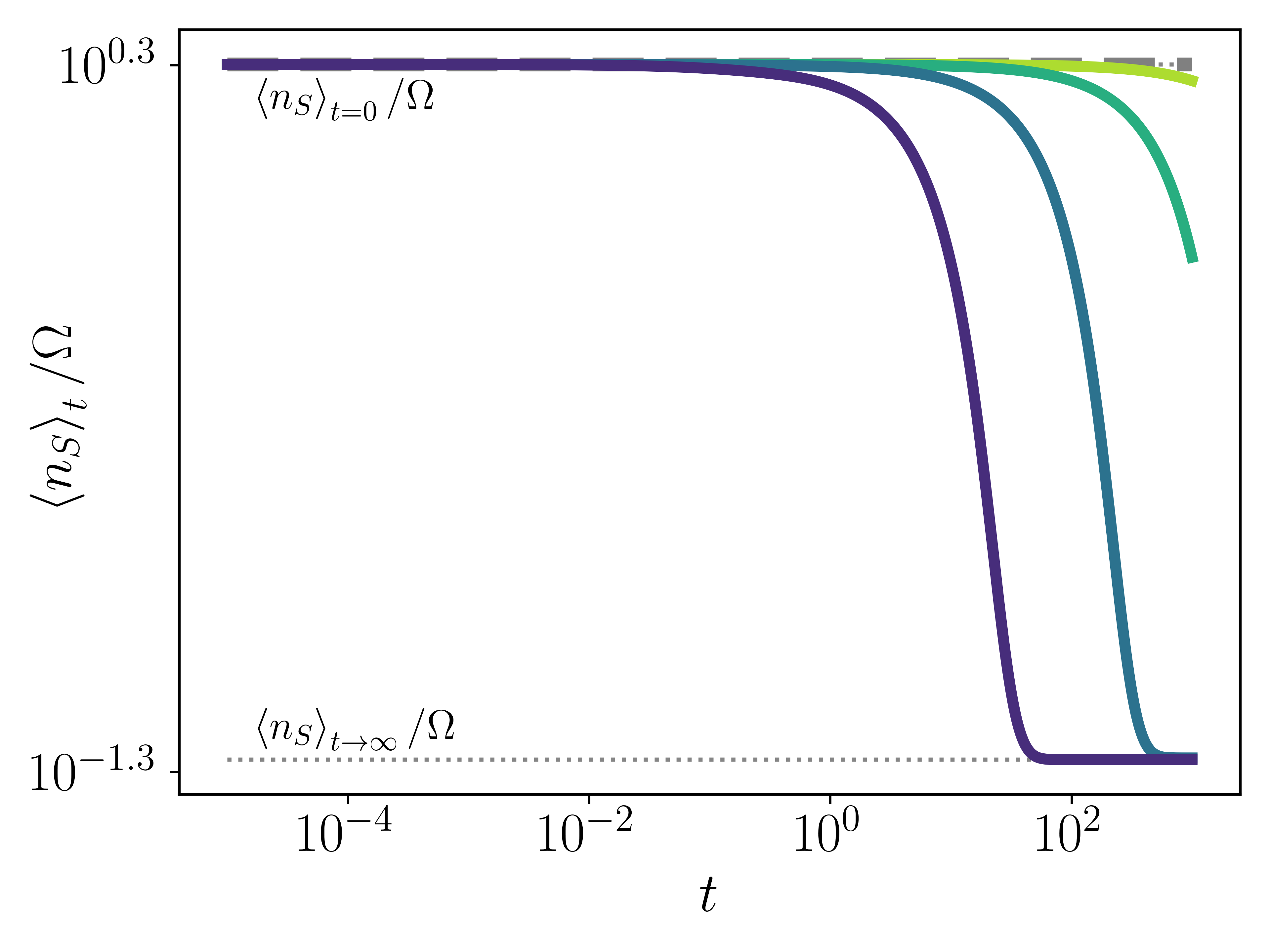}}
	\subfloat[\label{sfig:1c}]{%    
		\includegraphics[width=0.32\textwidth]{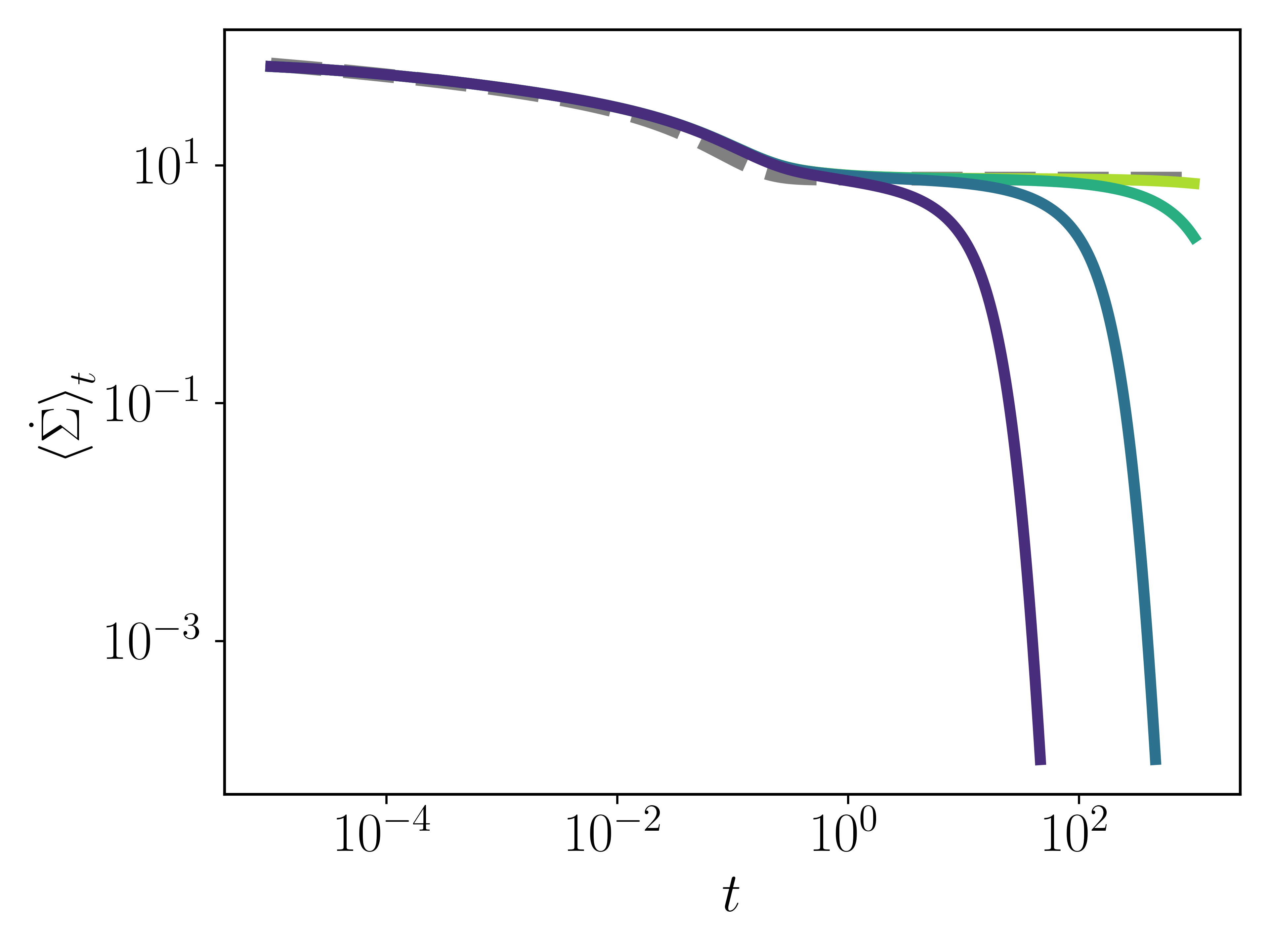}}
	\caption{
    Dynamics and thermodynamics of the {\col Michaelis}-Menten CRN~\eqref{eq:MMscheme}.
    (a) Mean molecule number of the enzyme~$E$
    (with
    $\braket{n_E}_{t=0}$, 
    $\braket{n_E}_{t\to\infty}$, 
    and $\braket{n_E}^{(0)}_{t\to\infty}$
    denoting 
    its initial value,
    its long-time value resulting from the CME~\eqref{eq:CME} 
    for the closed {\col Michaelis}-Menten CRN,
    and its long-time value resulting from the CME~\eqref{eq:CMElot} 
    for the hybrid {\col Michaelis}-Menten CRN,
    respectively). 
    (b) Mean concentration of the substrate~$S$
    (with
    $\braket{n_S}_{t=0}$
    and
    $\braket{n_S}_{t\to\infty}$
    denoting 
    the initial mean molecule number
    and
    its long-time value resulting from the CME~\eqref{eq:CME} 
    for the closed {\col Michaelis}-Menten CRN, 
    respectively). 
    (c) Mean entropy production rate~$\braket{\dot \Sigma}_t$. 
    % -
    The initial distribution of the closed {\col Michaelis}-Menten CRN is
    $P_{t=0}(n_E,n_C,n_S,n_P) =\delta(n_E-1)\delta(n_C)\delta(n_S - [S]_0\Omega)\delta(n_P - [P]_0\Omega)$,
    the initial distribution of the hybrid {\col Michaelis}-Menten CRN is
    $p_{t=0}(n_E,n_C) =\delta(n_E-1)\delta(n_C)$,
    with the initial concentrations (in arbitrary units) $[S]_0 = 2$ and $[P]_0 = 1$.
    The rate constants are (in arbitrary units) {\col $\log k_{+1} = \log k_{+2} = -\log k_{-1} = -\log k_{-2} = 1$}.}
	\label{fig:1}
\end{figure*}
    
    We consider the Michaelis-Menten CRN
    \begin{equation}
        \begin{split}
           S + E &\ch{<=>[$1$][$-1$]} C \,,\\
           C &\ch{<=>[$2$][$-2$]} E + P\,, 
        \end{split}
    \label{eq:MMscheme}
    \end{equation}
    representing the interconversion of a substrate~$S$ into a product~$P$ 
    catalyzed by the enzyme~$E$
    via an intermediate complex~$C$. 
    % -
    According to mass-action kinetics,
    the reaction rates of the Michaelis-Menten CRN read
    \begin{subequations}
        \begin{align}
            R_{1}(n_E,n_S) &= k_{1} n_E n_S/\Omega \,,\\
            R_{-1}(n_C) &= k_{-1} n_C \,,\\
            R_{2}(n_C) &= k_{2} n_C \,,\\
            R_{-2}(n_E,n_P) &= k_{-2} n_E n_P/\Omega \,.
        \end{align}
        \label{eq:MMReactionRatesFull}
    \end{subequations}
    % -

    % -
    We focus here on the case where
    substrate and product are high-abundant species,
    $[\boldsymbol{y}] \equiv ([S],[P])$,
    while enzyme and complex are low-abundant species,
    $\boldsymbol{n}_x \equiv (n_E,n_C)$.
    % -
    % -
    The corresponding hybrid Michaelis-Menten CRN has only discrete reactions, 
    i.e.,
    none of the reactions in Eq.~\eqref{eq:MMscheme} 
    interconverts directly the high-abundant species, namely, substrate and product,
    without interconverting the low-abundant species, namely, enzyme and complex.
    % -
    The corresponding reaction rates~\eqref{eq:MMReactionRatesFull}
    satisfy the scaling condition in Eq.~\eqref{eq:ReactionRatesScaling}
    and, 
    to leading order of the partial macroscopic limit,
    converge to
    \begin{subequations}
        \begin{align}
            r_{1}(n_E,[S]) &\simeq R_{1}(n_E| [S]_0) = k_{1} n_E [S]_0 \,,\\
            r_{-1}(n_C) &= R_{-1}(n_C) = k_{-1} n_C \,,\\
            r_{2}(n_C) &= R_{2}(n_C) = k_{2} n_C \,,\\
            r_{-2}(n_E,[P]) &\simeq R_{-2}(n_E | [P]_0) = k_{-2} n_E [P]_0 \,,
        \end{align}
        \label{eq:MMReactionRatesHybrid}%
    \end{subequations}
    where $[S]_0$ and $[P]_0$ denote 
    the initial concentration of the substrate and product,
    which do not evolve in time in the hybrid description 
    since there are no continuous reactions.
    % -
    % -

    % -
    We now compare the dynamics and thermodynamics of 
    the closed and hybrid 
    Michaelis-Menten CRNs 
    for different volumes $\Omega$,
    but with the same
    concentrations $([S]_0,[P]_0)$ 
    and total molecular number $n_E +n_C$.
    % -
    To do so, we numerically solve the CMEs~\eqref{eq:CME} and~\eqref{eq:CMElot}
    with the reaction rates 
    in Eq.~\eqref{eq:MMReactionRatesFull} and~\eqref{eq:MMReactionRatesHybrid},
    respectively.
    % -

    % -
    The dynamics of 
    the closed Michaelis-Menten CRN and the hybrid one 
    converge on short time scales.
    % -
    % -
    Figure~\ref{sfig:1a} shows the convergence of 
    the mean molecule number of the enzyme $\braket{n_E}_t$, 
    while
    Fig.~\ref{sfig:1b} shows the convergence of 
    the mean concentration of the substrate $\braket{n_S}_t/\Omega$ 
    of the closed Michaelis-Menten CRN
    to
    the constant concentration $[S]_0$ of the hybrid Michaelis-Menten CRN.
    % -
    By increasing the volume~$\Omega$,
    the convergence between the two descriptions 
    holds for longer time scales,
    but always fails in the long time limit 
    since the hybrid description does not capture 
    the evolution of the high-abundant species 
    (which play the role of autonomous chemostats).
    % -
    % -
    Note that the non-monotonic behavior of $\braket{n_E}_t$ 
    is just a consequence of the specific initial conditions used in this case
    (as already discussed in, for instance, Ref.~\onlinecite{nico20}).
    % -

    % -
    The same behavior is observed at the thermodynamic level too.
    % -
    Figure~\ref{sfig:1c} shows that
    the mean entropy production rates $\braket{\dot{\Sigma}}_t$ 
    computed with the two descriptions 
    converge on short time scales 
    (which can be increased by increasing the volume $\Omega$)
    and
    diverge in the long time limit.
    % -
    % -
    Indeed, the closed Michaelis-Menten CRN relaxes eventually to equilibrium
    characterized by $\braket{\dot{\Sigma}}_t = 0$,
    while the hybrid Michaelis-Menten CRN stays out of equilibrium 
    because of the high-abundant species acting as autonomous chemostats.
    % -

	\subsection{Cyclic Michaelis-Menten}

    \begin{figure*}
	\subfloat[\label{sfig:2a}]{%
		\includegraphics[width=0.33\textwidth]{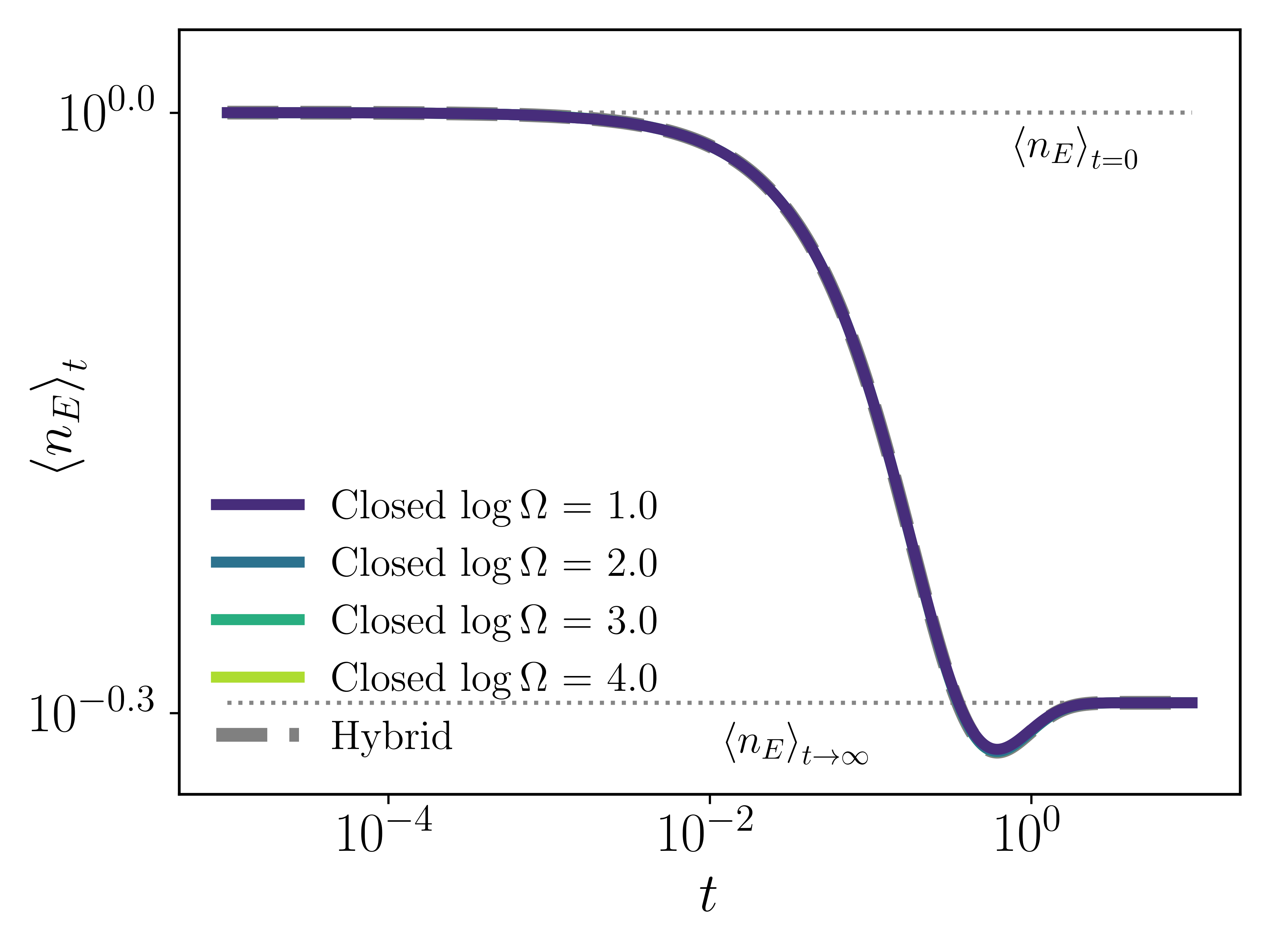}}     
	\subfloat[\label{sfig:2b}]{%
    \includegraphics[width=0.33\textwidth]{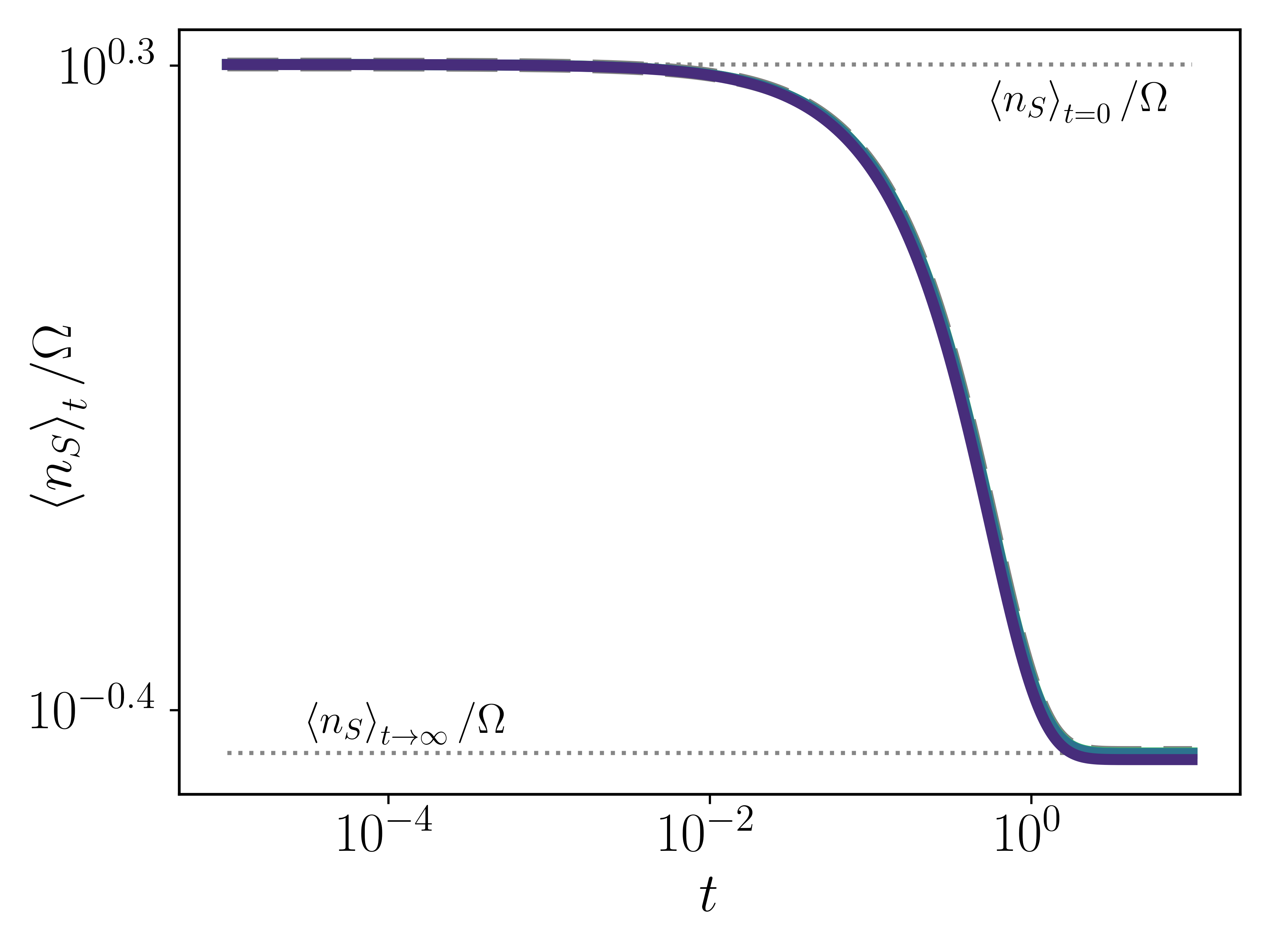}}

    \subfloat[\label{sfig:2c}]{%
		\includegraphics[width=0.33\textwidth]{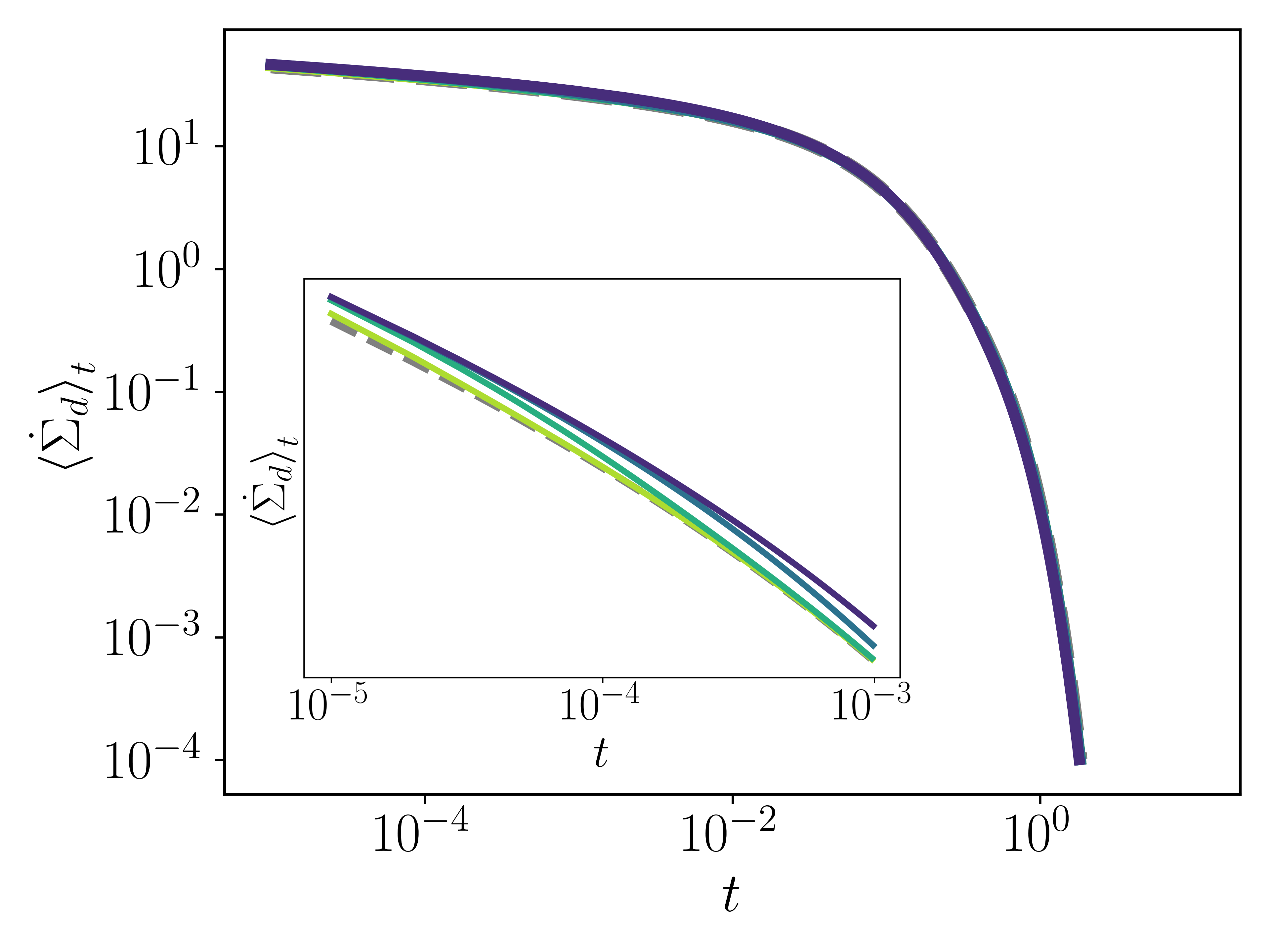}}
    \subfloat[\label{sfig:2d}]{%    
		\includegraphics[width=0.33\textwidth]{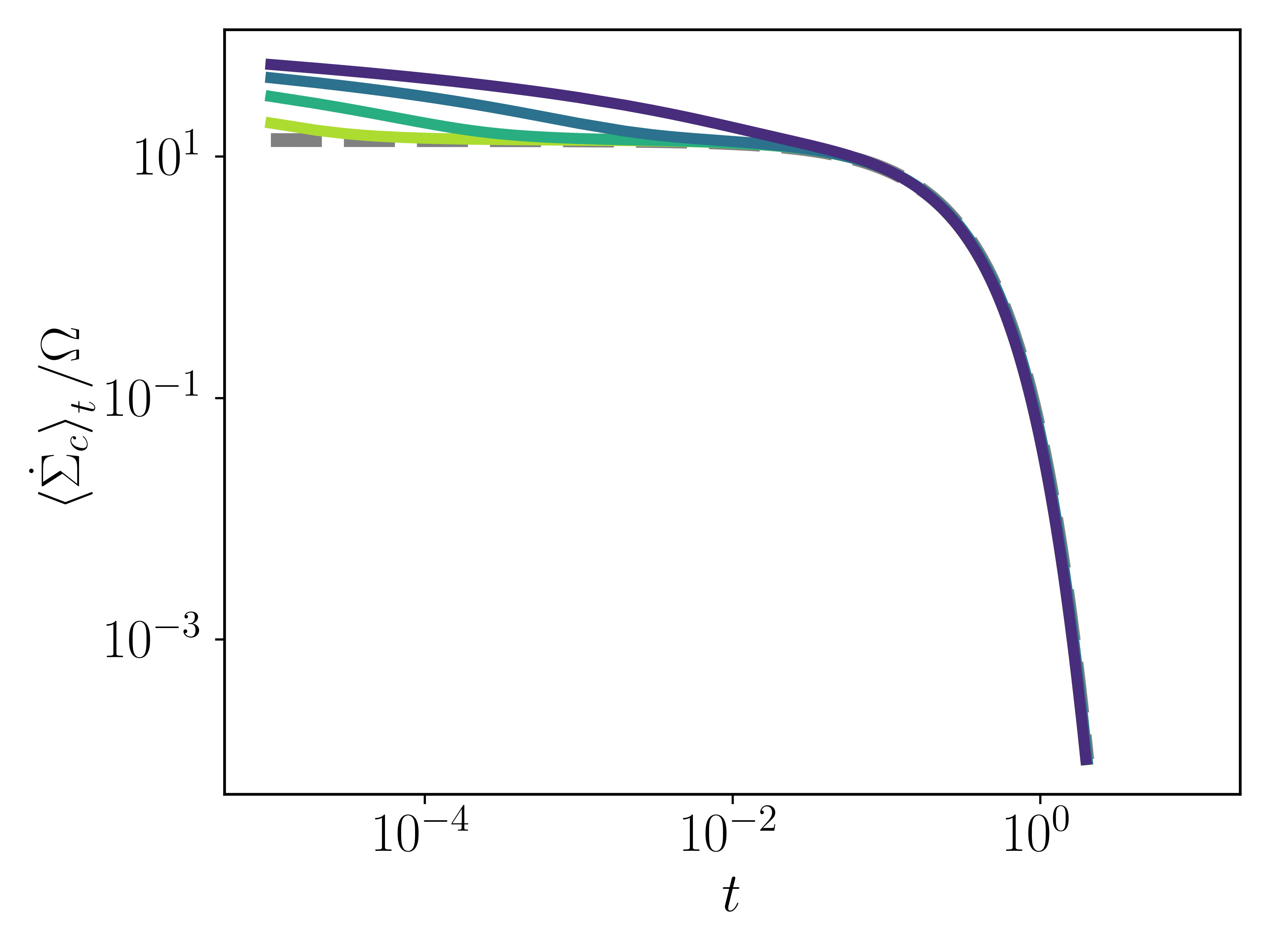}}
	\caption{
    Dynamics and thermodynamics of the cyclic {\col Michaelis}-Menten CRN~\eqref{eq:MMcycScheme}.
    (a) Mean molecule number of the enzyme~$E$
    (with
    $\braket{n_E}_{t=0}$
    and $\braket{n_E}_{t\to\infty}$, 
    denoting 
    its initial value
    and its long-time value resulting from the CME~\eqref{eq:CME} 
    for the closed cyclic {\col Michaelis}-Menten CRN,
    respectively). 
    (b) Mean concentration of the substrate~$S$
    (with
    $\braket{n_S}_{t=0}$
    and
    $\braket{n_S}_{t\to\infty}$
    denoting 
    the initial mean molecule number
    and
    its long-time value resulting from the CME~\eqref{eq:CME} 
    for the closed cyclic {\col Michaelis}-Menten CRN, 
    respectively). 
    (c) Mean entropy production rate of the discrete reactions $\braket{\dot \Sigma_d}_t$. 
    (d) Mean entropy production rate of the continuous reaction $\braket{\dot \Sigma_c}_t$. 
    % -
    The initial distribution of the closed cyclic {\col Michaelis}-Menten CRN is
    $P_{t=0}(n_E,n_C,n_S,n_P) =\delta(n_E-1)\delta(n_C)\delta(n_S - [S]_0\Omega)\delta(n_P - [P]_0\Omega)$,
    the initial distribution of the hybrid cyclic {\col Michaelis}-Menten CRN is
    $p_{t=0}(n_E,n_C) =\delta(n_E-1)\delta(n_C)$,
    with the initial concentrations (in arbitrary units) $[S]_0 = 2$ and $[P]_0 = 1$.
    The rate constants are (in arbitrary units) $\log k_{1} = \log k_{2} = -\log k_{-1} = -\log k_{-2} = 1/2$ and $\log k_3 = - \log k_{-3} = 1$.}
	\label{fig:2}
\end{figure*}

    We consider the cyclic Michaelis-Menten CRN~\cite{mare24b}
    \begin{equation}
    \begin{split}
        S + E & \ch{<=>[$1$][$-1$]} C \,, \\
        C & \ch{<=>[$2$][$-2$]} E + P \,, \\
        S & \ch{<=>[$3$][$-3$]} P \,,
    \end{split}
    \label{eq:MMcycScheme}
    \end{equation}
    where 
    the reactions $\pm1$ and $\pm2$ are the same catalytic reactions 
    of Eq.~\eqref{eq:MMscheme},
    while reactions $\pm3$ represent the direct interconversion
    of the substrate into the product.
    % -
    According to mass-action kinetics, 
    the rates of reactions $\pm1$ and $\pm2$ are given in Eq.~\eqref{eq:MMReactionRatesFull},
    while the rates of reactions $\pm3$ read
    \begin{subequations}
        \begin{align}
            R_{3}(n_S) &= k_{3} n_S \,,\\
            R_{-3}(n_P) &= k_{-3} n_P \,.
        \end{align}
    \end{subequations}
    % -
    Note that thermodynamic consistency, 
    i.e., the local detailed balance condition~\eqref{eq:LDBopen},
    imposes 
    that the kinetic constants of the reactions in Eq.~\eqref{eq:MMcycScheme} satisfy
    \begin{equation}
        \frac{k_{1}k_{2}k_{-3}}
        {k_{-1}k_{-2}k_{3}}
        = 1
        \,.
        \label{eq:MMCycLDB}
    \end{equation}
    This condition, also known as Wegscheider’s condition,~\cite{Schuster1989}
    ensures that the closed cyclic Michaelis-Menten CRN equilibrates.
    % -

    % -
    As in Subs.~\ref{sec:MM},
    we focus here on the case where 
    substrate and product are high-abundant species.
    % - 
    In the corresponding hybrid description,
    reactions $\pm3$ are continuous reactions
    determining the evolution of the concentrations $[\boldsymbol y]_t =([S]_t, [P]_t)$
    according to Eq.~\eqref{eq:CMassActionevolve}
    with the rates given, to leading order of the partial macroscopic limit, by 
    \begin{subequations}
        \begin{align}
            r_{3}([S]_t) &\simeq R_{3}([S]_t)/\Omega = k_{3} [S]_t \,,\\
            r_{-3}([P]_t) &\simeq R_{-3}([P]_t)/\Omega = k_{-3} [P]_t\,.
        \end{align}
    \end{subequations}
    % -
    The resulting dynamics for $([S]_t, [P]_t)$ can be analytically determined 
    and reads
    \begin{equation}\small
        [S]_t = 
        [S]_0 e^{-(k_3 + k_{-3})t}
        + \frac{k_{-3}([S]_0 + [P]_0)}{k_3 + k_{-3}}
        \bigg(1 - e^{-(k_3 + k_{-3})t}\bigg)
        \label{eq:MMcycSbar}
    \end{equation}
    and $[P]_t = ([S]_0 + [P]_0) - [S]_t$.
    % -
    
    % -
    Reactions $\pm 1$ and $\pm 2$ are still discrete reactions,
    whose rates are still given, to leading order,
    in Eq.~\eqref{eq:MMReactionRatesHybrid}
    once the initial concentrations $([S]_0, [P]_0)$ 
    are replaced by the instantaneous ones $([S]_t, [P]_t)$.
    % -
    
    % -
    We now compare the dynamics and thermodynamics of 
    the closed and hybrid 
    cyclic Michaelis-Menten CRNs 
    for different volumes $\Omega$,
    but with the same
    initial concentrations $([S]_0,[P]_0)$ 
    and total molecular number $n_E +n_C$.
    % -
    To do so, we numerically solve the CMEs~\eqref{eq:CME} and~\eqref{eq:CMElot}
    with the concentrations $[\boldsymbol y]_t$ given in Eq.~\eqref{eq:MMcycSbar}.
    % -

    % -
    The dynamics of 
    the closed cyclic Michaelis-Menten CRN and the hybrid one 
    converge on both short and long time scales.
    % -
    % -
    Figure~\ref{sfig:2a} shows the convergence of 
    the mean molecule number of the enzyme $\braket{n_E}_t$, 
    while
    Fig.~\ref{sfig:2b} shows the convergence of 
    the mean concentration of the substrate $\braket{n_S}_t/\Omega$ 
    of the closed cyclic Michaelis-Menten CRN
    to
    the concentration $[S]_t$ of the hybrid cyclic Michaelis-Menten CRN.
    % -
    % -
    The convergence on long time scales results from the continuous reaction
    which ensures that 
    the evolution of the high-abundant species in the hybrid description
    well reproduces their evolution in the closed cyclic Michaelis-Menten CRN.
    % -
    % -
    Note that, like in Fig.~\ref{sfig:1a},
    the non-monotonic behavior of $\braket{n_E}_t$ 
    is just a consequence of the specific initial conditions used in this case
    (as already discussed in, for instance, Ref.~\onlinecite{nico20}).
    % -

    A similar behavior, 
    namely, a convergence of the two descriptions at both short and long time scales,
    is observed for the mean entropy production rates of the discrete reactions 
    in Fig.~\ref{sfig:2c}. 
    % -
    Furthermore, 
    in both descriptions, the CRN relaxes eventually to equilibrium 
    characterized by $\braket{\dot{\Sigma}_d}_t = 0$
    and, indeed, the  mean entropy production rates in Fig.~\ref{sfig:2c} 
    monotonously decrease.
    % -
    On the other hand,
    the mean entropy production rates of the continuous reaction
    $\braket{\dot{\Sigma}_c}_t$ in Fig.~\ref{sfig:2d}
    computed with the two descriptions 
    converge on long time scales 
    (where the reaction equilibrates and $\braket{\dot{\Sigma}_c}_t = 0$), but
    do not match on short time scales for small volumes $\Omega$.
    % - 
    This is due to
    the correlations between the low- and high-abundant species 
    in the closed cyclic Michaelis-Menten CRN,
    which are lost in the hybrid cyclic Michaelis-Menten CRN 
    where the evolution of the high-abundant species is independent 
    of the low-abundant ones.
    % -
    These correlations become less relevant for larger volumes $\Omega$
    (see Fig.~\ref{sfig:2d}).
    % -
    Note that a similar trend can be observed for the 
    mean entropy production rates of the discrete reactions 
    in the inset in Fig.~\ref{sfig:2c}.
    % -
    However, the mismatch for $\braket{\dot{\Sigma}_d}_t$ on short time scales 
    is of a smaller magnitude compared to
    the mismatch for $\braket{\dot{\Sigma}_c}_t$
    since the discrete reactions depend on the high-abundant species 
    also in the hybrid cyclic Michaelis-Menten CRN 
    and hence keep track of the correlations between the two sets of species.

%%%%%%%%%%%%%%%%%%%%%%%%%%%%%%%%%%%%%%%%%%%%%%%%%%%%%%%%%%%%
%%%%%%%%%%%%%%%%%%%%%%%%%%%%%%%%%%%%%%%%%%%%%%%%%%%%%%%%%%%%
	% \newpage\text{}
    % \newpage
	\section{Discussion \& Conclusion}
    \label{sec:Discussion}

    \begin{figure}
	\subfloat[\label{sfig:3a}]{%
		\includegraphics[width=0.33\textwidth]{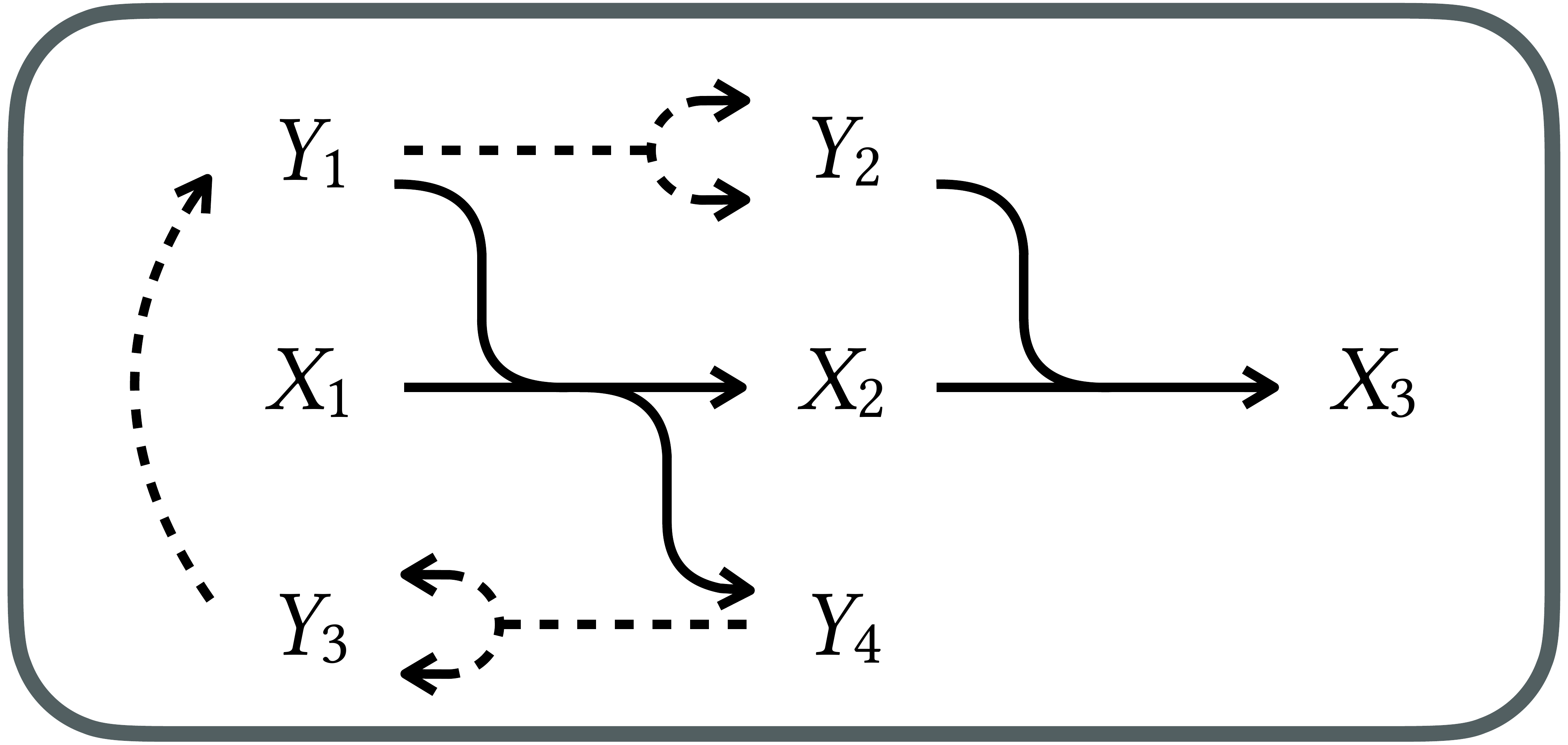}}   
        
	\subfloat[\label{sfig:3b}]{%
    \includegraphics[width=0.33\textwidth]{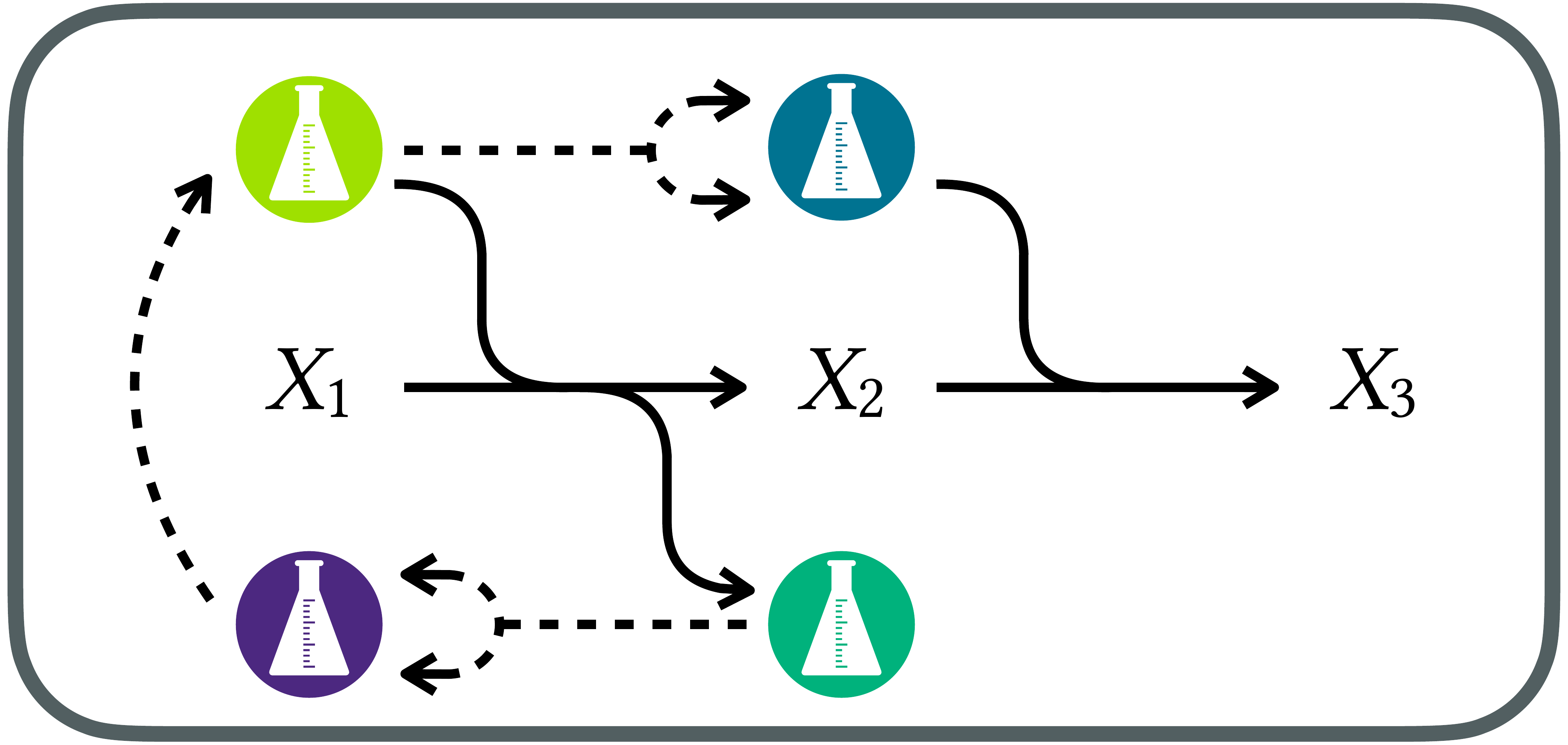}}
	\caption{
    Exemplary hybrid CRN whose stoichiometry satisfies 
    {\col 
    the conditions 
    in Eqs.~\eqref{eq:ConstraintDiscrete} and~\eqref{eq:ConstraintContinuous}}.
    % the scaling conditions in Eq.~\eqref{eq:ReactionRatesScaling}.
    % -
    % -
    (a)~Closed CRN with 
    low-abundant species $(X_1, X_2, X_3)$ and 
    high-abundant species $(Y_1, Y_2, Y_3, Y_4)$. 
    % -
    (b)~Open CRN with internal species $(X_1, X_2, X_3)$ and 
    chemostats represented by Erlenmeyer flasks of different colors. 
    % -
    % -
    In the partial macroscopic limit, 
    the closed CRN~(a) converges to the open one~(b). 
    % -
    % -
    Arrows represent the forward chemical reactions using hypergraph notation,
    while backward reactions are not illustrated for simplicity. 
    % -
    Solid arrows represent discrete reactions (Eq.~\eqref{eq:DefDiscreteReactions})
    that are unimolecular in the low-abundant (resp. internal) species,
    while having arbitrary stoichiometry in the high-abundant species (resp. chemostats).
    % -
    Dashed arrows represent continuous reactions (Eq.~\eqref{eq:DefContinuousReactions})
    that do not involve low-abundant (resp. internal) species
    and have arbitrary stoichiometry in the high-abundant species (resp. chemostats).
    }
	\label{fig:3}
\end{figure}

    In this work,
    we showed that the dynamics (Sec.~\ref{sec:HDele}) 
    and thermodynamics (Sec.~\ref{sec:TD})
    of open CRNs following mass-action kinetics
    can emerge from closed CRNs,
    when specific stoichiometric conditions are satisfied 
    (see Fig. \ref{fig:3}).
    % -
    In particular,
    open CRNs with unimolecular reactions in the internal species
    result from 
    the partial macroscopic limit of
    hybrid CRNs 
    with 
    discrete reactions that are unimolecular in the low-abundant species 
    (see Eq.~\eqref{eq:ConstraintDiscrete})
    and 
    continuous reactions that do not involve the low-abundant species
    (see Eq.~\eqref{eq:ConstraintContinuous}).
    % -
    % -
    On the one hand,
    the low-abundant species
    and the discrete reactions of hybrid CRNs
    become the internal species and the reactions of open CRNs, respectively.
    % -
    Correspondingly, 
    the entropy production rate of the discrete reactions quantifies 
    the dissipation of open CRNs.
    % -
    % -
    On the other hand,
    the high-abundant species
    and the continuous reactions of hybrid CRNs
    become the chemostats of open CRNs and their driving protocols, respectively.
    % -
    Correspondingly, 
    the entropy production rate of the continuous reactions quantifies 
    the dissipation of the driving protocols,
    usually neglected in the modeling of open CRNs.
    % -
    % -
    
    % -
    We note that, to the best of our knowledge, 
    only Ref.~\onlinecite{fagg09} examined thermodynamic-like observables for hybrid CRNs
    by establishing 
    a fluctuation relation 
    for the dynamics of the high-abundant species.
    % -
    However, contrary to our work,
    they built this relation directly on the hybrid dynamics 
    instead of considering how it emerges from the microscopic scale 
    by taking the partial macroscopic limit.
    % -
        
    % -
    % -
    While our work clearly demonstrates how chemostats can result
    from an underlying closed CRN,
    it does so for a subset of all possible stoichiometries. 
    % -
    This {\col limitation arises from} 
    % is a direct implication of 
    the {\col partial macroscopic limit}
    % large volume limit 
    developed in Refs.~\onlinecite{ball06,menz12,crud12,kang13,ande15,wink17,huft19}
    which is based on a single {\col time}-scale perturbation analysis
    {\col(as discussed at the end of subs.~\ref{sec:HybridGeneral})},
    % -
    {\col 
    while multi-molecular reactions evolve over multiple time-scales.
    Hence, }
    to investigate
    the emergence of chemostats in CRNs not satisfying the stoichiometric conditions in Eqs.~\eqref{eq:ConstraintDiscrete} and~\eqref{eq:ConstraintContinuous}, 
    a multiple {\col time}-scale perturbation analysis~\cite{kuehn2015}
    needs to be 
    {\col first} developed and {\col then} used. 
    We leave this to further studies.    

    % -
    \begin{acknowledgments}
    BR and ME are supported by the Fond National de la Recherche-FNR, Luxembourg
    by the Project ChemComplex (C21/MS/16356329).
    % -
    FA is supported by the project P-DiSC\#BIRD2023-UNIPD 
    funded by the Department of Chemical Sciences of the University of Padova (Italy).

    \end{acknowledgments}

%%%%%%%%%%%%%%%%%%%%%%%%%%%%%%%%%%%%%%%%%%%%%%%%%%%%%%%%%%%%
%%%%%%%%%%%%%%%%%%%%%%%%%%%%%%%%%%%%%%%%%%%%%%%%%%%%%%%%%%%%

	\section*{Data Availability}
    The data that support the findings of this study are available from the corresponding author upon reasonable request.

%%%%%%%%%%%%%%%%%%%%%%%%%%%%%%%%%%%%%%%%%%%%%%%%%%%%%%%%%%%%
%%%%%%%%%%%%%%%%%%%%%%%%%%%%%%%%%%%%%%%%%%%%%%%%%%%%%%%%%%%%

\onecolumngrid
\appendix

\section{Derivation of the dynamics of hybrid CRNs}
\label{appendix:hybrid}

We review here the general derivation of the hybrid dynamics 
in the partial macroscopic limit
summarized in Subs.~\ref{sec:HybridGeneral}
by following the original derivation in Ref.~\onlinecite{menz12}
(which does not assume that the reaction rates satisfy mass-action kinetics).
We start by 
rewriting the CME~\eqref{eq:CME} for hybrid CRNs (defined in Subs.~\ref{sec:hCRNsetup})
as 
   \begin{equation}
        \begin{split}
            \partial_t P_t(\boldsymbol n_x,\boldsymbol n_y) = 
            &\sum\limits_{\rho \in \mathcal R_d} 
            \big\{
            R_\rho(\boldsymbol{n}_x-\mathbf S_{x,\rho},\boldsymbol{n}_y-\mathbf S_{y,\rho})
            P_t(\boldsymbol{n}_x-\mathbf S_{x,\rho},\boldsymbol{n}_y-\mathbf S_{y,\rho}) 
            -R_\rho(\boldsymbol{n}_x,\boldsymbol{n}_y)
            P_t(\boldsymbol{n}_x,\boldsymbol{n}_y)
            \big\}\\
            &+\sum\limits_{\rho \in \mathcal R_c} 
            \big\{
            R_\rho(\boldsymbol{n}_x,\boldsymbol{n}_y-\mathbf S_{y,\rho})
            P_t(\boldsymbol{n}_x,\boldsymbol{n}_y-\mathbf S_{y,\rho}) 
            -R_\rho(\boldsymbol{n}_x,\boldsymbol{n}_y)
            P_t(\boldsymbol{n}_x,\boldsymbol{n}_y)
            \big\}
            \,,
        \end{split}
        \label{eq:FullCME}
    \end{equation}
by accounting for
the splitting of the chemical species into 
low-abundant (labeled~$x$)
and high-abundant (labeled~$y$) species
as well as
the splitting of the chemical reactions $\mathcal R$ into 
discrete $\mathcal R_d$
and continuous $\mathcal R_c$ reactions.
% -
% -
We then take the partial macroscopic limit,
namely, 
the large volume limit $\Omega \to \infty$ 
characterized by
finite molecule numbers $\boldsymbol{n}_x$
and
finite concentrations $[\boldsymbol{y}]\equiv(\dots, [y] ,\dots)$,
of the CME~\eqref{eq:FullCME}.
% -
To do so, we need to consider how 
the reaction rates $\{R_\rho(\boldsymbol n_x, \boldsymbol n_y)\}$ and 
the probability $P_t(\boldsymbol n_x, \boldsymbol n_y)$ depend on $\Omega$.
% -

% -
First, we consider the reaction rates $\{R_\rho(\boldsymbol n_x, \boldsymbol n_y)\}$.
By assuming the scaling conditions in 
Eqs.~\eqref{eq:def_rates_hCRNs} and~\eqref{eq:ReactionRatesScaling},
the CME~\eqref{eq:FullCME} becomes
    \begin{equation}
        \begin{split}
            \partial_t P_t(\boldsymbol n_x,\boldsymbol n_y) = 
            &\sum\limits_{\rho \in \mathcal R_d} 
            \bigg\{
            r_\rho\bigg(\boldsymbol{n}_x-\mathbf S_{x,\rho},[\boldsymbol y]-\frac{\mathbf S_{y,\rho}}{\Omega}\bigg)
            P_t(\boldsymbol{n}_x-\mathbf S_{x,\rho},\boldsymbol{n}_y-\mathbf S_{y,\rho}) 
            -r_\rho(\boldsymbol{n}_x,[\boldsymbol y])
            P_t(\boldsymbol{n}_x,\boldsymbol{n}_y)
            \bigg\}\\
            &+\Omega\sum\limits_{\rho \in \mathcal R_c} 
            \bigg\{
            r_\rho\bigg(\boldsymbol{n}_x,[\boldsymbol y]-\frac{\mathbf S_{y,\rho}}{\Omega}\bigg)
            P_t(\boldsymbol{n}_x,\boldsymbol{n}_y-\mathbf S_{y,\rho}) 
            -r_\rho(\boldsymbol{n}_x,[\boldsymbol y])
            P_t(\boldsymbol{n}_x,\boldsymbol{n}_y)
            \bigg\}
            \,,
        \end{split}
        \label{eq:FullCME2}
    \end{equation}
with $r_\rho(\boldsymbol n_x,[\boldsymbol{y}]) = \mathcal O(1)$
for  $\Omega \to \infty$.
% -
% -

% -
Second, we consider the probability $P_t(\boldsymbol n_x, \boldsymbol n_y)$.
We rewrite it as
\begin{equation}
    P_t(\boldsymbol{n}_x,\boldsymbol{n}_y) = P_t(\boldsymbol{n}_y|\boldsymbol{n}_x)P_t(\boldsymbol{n}_x)
    \,,
\end{equation}
and use the change of variables from $\boldsymbol{n}_y$ to $[\boldsymbol{y}]$ to define
\begin{equation}
    p^{(\Omega)}_t([\boldsymbol{y}]|\boldsymbol{n}_x)
    p^{(\Omega)}_t(\boldsymbol{n}_x) 
    \equiv 
    \Omega^{N_y} P_t(\Omega[\boldsymbol y]|\boldsymbol{n}_x)
    P_t(\boldsymbol{n}_x)
    \,,
    \label{eq:prob_def_scaling}
\end{equation}
where $N_y$ denotes the total number of high-abundant species
and the superscript $\Omega$ highlights that the probabilities 
on the left-hand side of Eq.~\eqref{eq:prob_def_scaling} 
depend on the volume.
% -
In the partial macroscopic limit,
we use the Wentzel–Kramers–Brillouin (WKB) ansatz~\cite{risken,touc09,bres14,touc18,qian21,fala25} 
for the conditional probability distribution
$p^{(\Omega)}_t([\boldsymbol{y}]|\boldsymbol{n}_x)$:
\begin{equation}
		p^{(\Omega)}_t([\boldsymbol{y}]|\boldsymbol{n}_x) 
        = C_\Omega 
        e^{ -\Omega I_t([\boldsymbol{y}]|\boldsymbol{n}_x)}
        \sum_{k=0}^\infty  U^{(k)}_t([\boldsymbol{y}]|\boldsymbol{n}_x) \Omega^{-k}
        \,,
    \label{eq:WKB}
    \end{equation}
where 
$C_\Omega$ is the normalization constant,
$I_t([\boldsymbol{y}]|\boldsymbol{n}_x)$ is the zero-order rate function,
and each $U^{(k)}_t([\boldsymbol{y}]|\boldsymbol{n}_x)$ is 
a subexponential correction of order $\Omega^{-k}$.
% -
Furthermore, 
we use a perturbation expansion series 
for the marginalized distribution of the discrete species
$p^{(\Omega)}_t(\boldsymbol{n}_x)$
in powers of $1/\Omega$:
	\begin{equation}
		p^{(\Omega)}_t(\boldsymbol{n}_x) 
        =
        p_t(\boldsymbol{n}_x) 
        +
        \sum_{k=1}^\infty p^{(k)}_t(\boldsymbol{n}_x)\Omega^{-k}
        \,.
        \label{eq:PerturbationAnsatz}
	\end{equation}   
By now using Eqs.~\eqref{eq:WKB} and~\eqref{eq:PerturbationAnsatz} 
in the CME~\eqref{eq:FullCME2},
the leading order $\mathcal O(\Omega)$ of its left-hand-side reads
\begin{equation}
-\Omega
C_\Omega 
e^{-\Omega I_t([\boldsymbol{y}] |\boldsymbol{n}_x)} U^{(0)}_t([\boldsymbol{y}]|\boldsymbol{n}_x) 
p_t(\boldsymbol{n}_x) 
\partial_t I_t([\boldsymbol y]|\boldsymbol n_x) 
% + \mathcal O(1)
\,,
\label{eq:LHS}
\end{equation}
while the leading order $\mathcal O(\Omega)$ of its right-hand-side
emerges only from the continuous reactions
and reads
\begin{equation}
    \Omega 
    C_\Omega 
    e^{-\Omega I_t([\boldsymbol{y}] |\boldsymbol{n}_x) }
    U^{(0)}_t([\boldsymbol{y}] |\boldsymbol{n}_x)
    p_t(\boldsymbol{n}_x)
    \sum\limits_{\rho \in \mathcal R_c}
    r_{\rho}(\boldsymbol{n}_x,[\boldsymbol{y}])
    \Big\{
    e^{ \nabla_{[\boldsymbol y]} I_t([\boldsymbol{y}] |\boldsymbol{n}_x) 
    \cdot \mathbf S_{y,\rho} }
    -1
    \Big\}
    \,,
    \label{eq:RHS}
\end{equation}
with $\nabla_{[\boldsymbol y]} = (\dots\,, \partial_{[y]}\,, \ldots)$.
% -
% -
Hence, the leading order $\mathcal O(\Omega)$ of the CME~\eqref{eq:FullCME2} reads
\begin{equation}
    \partial_t I_t([\boldsymbol y]|\boldsymbol n_x)
    =
    -
    \sum\limits_{\rho \in \mathcal R_c}
    r_{\rho}(\boldsymbol{n}_x,[\boldsymbol{y}])
    \Big\{
    e^{ \nabla_{[\boldsymbol y]} I_t([\boldsymbol{y}] |\boldsymbol{n}_x) 
    \cdot \mathbf S_{y,\rho} }
    -1
    \Big\}
    \equiv 
    - 
    H(
    [\boldsymbol y], 
    \nabla_{[\boldsymbol y]} I_t([\boldsymbol y]|\boldsymbol{n}_x)
    |\boldsymbol{n}_x)\,,
\end{equation}
where we introduced the Hamiltonian $H(\boldsymbol{q},\boldsymbol{\xi}|\boldsymbol{n}_x)$
with the variables $\boldsymbol{q}$ and the conjugated variables $\boldsymbol{\xi}$
corresponding to $[\boldsymbol y]$ and 
$\nabla_{[\boldsymbol y]} I_t([\boldsymbol y]|\boldsymbol{n}_x)$,
respectively.
% -

% -
Finally, we assume that the conditional probability distribution
$p^{(\Omega)}_t([\boldsymbol{y}]|\boldsymbol{n}_x)$ 
has a unique maximum at every time $t$,
denoted $[\boldsymbol{y}|\boldsymbol{n}_x]_t$,
with the necessary condition for its existence given by
\begin{equation}
    \boldsymbol{\xi}_t \equiv  
    \nabla_{[\boldsymbol y]} 
    I_t([\boldsymbol{y}]|\boldsymbol{n}_x)
    \big|_{[\boldsymbol{y}] = [\boldsymbol{y}|\boldsymbol{n}_x]_t} 
    \overset{!}{=} \boldsymbol 0
    \,.
    \label{eq:OptimalP}
\end{equation}
% -
% -
The existence of this maximum leads to 
the RRE~\eqref{eq:RRElot} for the concentrations of the high-abundant species
and the CME~\eqref{eq:CMElot} for the low-abundant species.
% -
Indeed, $[\boldsymbol{y}|\boldsymbol{n}_x]_t$ defines 
the most probable concentrations of the high-abundant species and
evolves in time according to the following Hamilton equation
\begin{equation}
    \partial_t [\boldsymbol{y}|\boldsymbol{n}_x]_t 
    = 
    \nabla_{\boldsymbol{\xi}}
    H([\boldsymbol{y}|\boldsymbol{n}_x]_t,\boldsymbol{\xi}|\boldsymbol{n}_x) 
    \big|_{\boldsymbol \xi = \boldsymbol 0}
    =
    \sum\limits_{\rho \in \mathcal R_c} 
    \mathbf S_{y,\rho} \, 
    r_\rho(\boldsymbol{n}_x,[\boldsymbol{y}|\boldsymbol{n}_x]_t)
    \,,
    \label{eq:condRRE}
    \end{equation}
which is RRE~\eqref{eq:RRElot} in the main text.
% -
% -
On the other hand, 
a CME for the low-abundant species can be obtained 
by marginalizing the CME~\eqref{eq:FullCME2},
\begin{equation}
    \begin{split}
    \partial_t p^{(\Omega)}_t(\boldsymbol{n}_x) 
    & = \int \mathrm d[\boldsymbol{y}]\,
        \partial_t \bigg\{
        p^{(\Omega)}_t([\boldsymbol y]|\boldsymbol{n}_x)
        p^{(\Omega)}_t(\boldsymbol{n}_x)
        \bigg\}\\
    & = \sum\limits_{\rho \in \mathcal R_d}
    \int \mathrm d[\boldsymbol{y}]\,
    \bigg\{
        r_\rho\bigg(\boldsymbol{n}_x-\mathbf S_{x,\rho},[\boldsymbol y]-\frac{\mathbf S_{y,\rho}}{\Omega}\bigg)
        p^{(\Omega)}_t\bigg([\boldsymbol y]-\frac{\mathbf S_{y,\rho}}{\Omega}\bigg|\boldsymbol{n}_x-\mathbf S_{x,\rho}\bigg)
        p^{(\Omega)}_t(\boldsymbol{n}_x-\mathbf S_{x,\rho})\\
        &\quad\quad\quad\quad\quad\quad\quad
        -r_\rho(\boldsymbol{n}_x, [\boldsymbol y])
        p^{(\Omega)}_t([\boldsymbol y]|\boldsymbol{n}_x)
        p^{(\Omega)}_t(\boldsymbol{n}_x)
    \bigg\}
    \,,
    \end{split}
    \label{eq:MarginalCME}
\end{equation}
together with
Laplace's saddle-node approximation, i.e.,
\begin{equation}
    \int\mathrm d[\boldsymbol{y}]\,
    p^{(\Omega)}_t([\boldsymbol y]|\boldsymbol{n}_x) f
    (\boldsymbol n_x,[\boldsymbol y]) 
    =
    f(\boldsymbol n_x,[\boldsymbol{y}|\boldsymbol{n}_x]_t)
    + 
    \mathcal{O}(\Omega^{-1})
    %{\color{blue}\mathcal{O}(\Omega^{-1/2})}    
    \,.
    \label{eq:LaplaceApprox}
\end{equation}
We, thus, find
\begin{equation}
    \begin{split}
        \partial_t p_t(\boldsymbol{n}_x) = 
        \sum\limits_{\rho \in \mathcal R_d} 
        \big\{
        r_\rho(\boldsymbol{n}_x-\mathbf S_{x,\rho},[\boldsymbol{y}|\boldsymbol{n}_x-\mathbf S_{x,\rho}]_t)p_t(\boldsymbol{n}_x-\mathbf S_{x,\rho})
         - r_\rho(\boldsymbol{n}_x,[\boldsymbol{y}|\boldsymbol{n}_x]_t)p_t(\boldsymbol{n}_x) \}
        \,.
        \label{eq:DiscreteCME}
    \end{split}
\end{equation}
which is CME~\eqref{eq:CMElot} in the main text.

%%%%%%%%%%%%%%%%%%%%%%%%%%%%%%%%%%%%%%%%%%%%%%%%%%%%%%%%%%%%
%%%%%%%%%%%%%%%%%%%%%%%%%%%%%%%%%%%%%%%%%%%%%%%%%%%%%%%%%%%%

\section{Derivation of the entropy production rate for hybrid CRNs}
\label{appendix:epr}

We derive here the partial macroscopic limit of the entropy production rate
for hybrid CRNs whose reactions satisfy the stoichiometric conditions in 
Eqs.~\eqref{eq:ConstraintDiscrete} and~\eqref{eq:ConstraintContinuous}.
% -
We start by rewriting its general expression in Eq.~\eqref{eq:EPRgeneral} as
\begin{equation}
    \begin{split}    
        \braket{\dot \Sigma}_t =
        & 
        \sum\limits_{\rho \in \mathcal R_d}
        \sum\limits_{\boldsymbol{n}_x,\boldsymbol{n}_y} 
        R_{\rho}(\boldsymbol{n}_x,\boldsymbol{n}_y)
        P_t(\boldsymbol{n}_x,\boldsymbol{n}_y)
        \ln \frac{
        R_{\rho}(\boldsymbol{n}_x,\boldsymbol{n}_y)
        P_t(\boldsymbol{n}_x,\boldsymbol{n}_y)
        }{
        R_{-\rho}(\boldsymbol{n}_x+\mathbf S_{x,\rho},\boldsymbol{n}_y+\mathbf S_{y,\rho})
        P_t(\boldsymbol{n}_x+\mathbf S_{x,\rho},\boldsymbol{n}_y+\mathbf S_{y,\rho})} \\
        & +
        \sum\limits_{\rho \in \mathcal R_c}
        \sum\limits_{\boldsymbol{n}_x,\boldsymbol{n}_y} 
        R_{\rho}(\boldsymbol{n}_y)
        P_t(\boldsymbol{n}_x,\boldsymbol{n}_y)
        \ln \frac{
        R_{\rho}(\boldsymbol{n}_y)
        P_t(\boldsymbol{n}_x,\boldsymbol{n}_y)
        }{
        R_{-\rho}(\boldsymbol{n}_y+\mathbf S_{y,\rho})
        P_t(\boldsymbol{n}_x,\boldsymbol{n}_y+\mathbf S_{y,\rho})}
    \end{split}
\end{equation}
by accounting for
the splitting of the chemical species into 
low-abundant (labeled~$x$)
and high-abundant (labeled~$y$) species
as well as
the splitting of the chemical reactions $\mathcal R$ into 
discrete $\mathcal R_d$
and continuous $\mathcal R_c$ reactions
(where the latter do not dependent on the low-abundant species because of 
stoichiometric conditions in Eq.~\eqref{eq:ConstraintContinuous}).
% -
% -
We then express 
the reaction rates in terms of $\{r_{\rho}(\boldsymbol n_x, [\boldsymbol y])\}$
(defined in Eq.~\eqref{eq:def_rates_hCRNs} and 
satisfying the scaling condition in Eq.~\eqref{eq:ReactionRatesScaling})
and 
the probability $P_t(\boldsymbol{n}_x,\boldsymbol{n}_x)$
in terms of $p^{(\Omega)}_t([\boldsymbol{y}]|\boldsymbol{n}_x)$ and 
$p^{(\Omega)}_t(\boldsymbol{n}_x)$ 
given in Eq.~\eqref{eq:prob_def_scaling}.
% -
We thus obtain
\begin{equation}\small
    \begin{split}
        \braket{\dot \Sigma}_t =
        & 
        \sum\limits_{\rho \in \mathcal R_d }
        \sum\limits_{\boldsymbol{n}_x} 
        \int \mathrm d[\boldsymbol{y}]\,
        r_{\rho}(\boldsymbol{n}_x,[\boldsymbol{y}])
        p^{(\Omega)}_t([\boldsymbol{y}] | \boldsymbol{n}_x)
        p^{(\Omega)}_t(\boldsymbol{n}_x)
        \ln \frac{
        r_{\rho}(\boldsymbol{n}_x,[\boldsymbol{y}])
        p^{(\Omega)}_t([\boldsymbol{y}]|\boldsymbol{n}_x)
        p^{(\Omega)}_t(\boldsymbol{n}_x)
        }{
        r_{-\rho}\Big(\boldsymbol{n}_x+\mathbf S_{x,\rho},[\boldsymbol{y}]+\frac{\mathbf S_{y,\rho}}{\Omega}\Big)
        p^{(\Omega)}_t\Big([\boldsymbol{y}]+\frac{\mathbf S_{y,\rho}}{\Omega}\Big|\boldsymbol{n}_x + \mathbf S_{x,\rho}\Big)
        p^{(\Omega)}_t(\boldsymbol{n}_x+\mathbf S_{x,\rho})}
        \\
        & +
        \Omega
        \sum\limits_{\rho \in \mathcal R_c}
        \sum\limits_{\boldsymbol{n}_x} 
        p^{(\Omega)}_t(\boldsymbol{n}_x)
        \int \mathrm d[\boldsymbol{y}] \,
        r_{\rho}([\boldsymbol{y}])
        p^{(\Omega)}_t([\boldsymbol{y}]|\boldsymbol n_x)
        \ln \frac{
        r_{\rho}([\boldsymbol{y}])
        p^{(\Omega)}_t([\boldsymbol{y}]|\boldsymbol n_x)
        }{
        r_{-\rho}\Big([\boldsymbol{y}]+\frac{\mathbf S_{y,\rho}}{\Omega} \Big)
        p^{(\Omega)}_t\Big([\boldsymbol{y}]+\frac{\mathbf S_{y,\rho}}{\Omega}\Big|\boldsymbol n_x\Big)}
        \,,
        \\
    \end{split}
\end{equation}
by replacing the sum $\sum_{\boldsymbol{n}_y}$ 
with the integral $\int \mathrm d[\boldsymbol{y}]$.
% -
% -
Finally,
by 
Taylor expanding the small terms $\mathbf S_{y,\rho}/\Omega$,
using Laplace's saddle-node approximation~\eqref{eq:LaplaceApprox},
and recognizing that the leading order contribution of
$p^{(\Omega)}_t([\boldsymbol{y}]|\boldsymbol n_x)$
as well as the corresponding maximum $[\boldsymbol y | \boldsymbol n_x]_t$
are $\boldsymbol n_x$-independent
if the continuous reactions 
do not involve the low-abundant species
(see Ref.~\onlinecite{menz12})
as imposed by the stoichiometric conditions~\eqref{eq:ConstraintContinuous},
the entropy production rate becomes, to leading order,
\begin{equation}
    \braket{\dot \Sigma}_t =
    \underbrace{\sum_{\rho\in \mathcal R_d,\boldsymbol{n}_x}
         r_{\rho}(\boldsymbol{n}_x,[\boldsymbol{y}]_t)
         p_t(\boldsymbol{n}_x)
         \ln \frac{
         r_{\rho}(\boldsymbol{n}_x,[\boldsymbol{y}]_t)
         p_t(\boldsymbol{n}_x)
         }{
         r_{-\rho}(\boldsymbol{n}_x+\mathbf S_{x,\rho},[\boldsymbol{y}]_t)p_t(\boldsymbol{n}_x+\mathbf S_{x,\rho})}
         }_{\simeq\braket{\dot{\Sigma_d}}_t}
         +
    \underbrace{\Omega
    \sum\limits_{\rho\in \mathcal R_c} 
        r_{\rho}([\boldsymbol{y}]_t)
        \ln 
        \frac{r_{\rho}([\boldsymbol{y}]_t)}{r_{-\rho}([\boldsymbol{y}]_t)}
        }_{\simeq\braket{\dot{\Sigma_c}}_t}
        \,,
\end{equation}
where we also recognize the entropy production rate
of the discrete and continuous reactions.
% -
% -
Note that the leading order contributions to the entropy production rate
of the discrete and continuous reactions are 
of order $\mathcal O(1)$ and $\mathcal O(\Omega)$, respectively.
% -
This implies that the next order contribution to $\braket{\dot{\Sigma}_c}_t$ 
is of the same order of $\braket{\dot{\Sigma}_d}_t$, i.e., $\mathcal O(1)$.
% -
This does not constitute an inconsistency of our theory since,
as explained in Subs.~\ref{sec:EPRcontinuous}, 
discrete and continuous reactions have different physical meanings.
The former represent the chemical reactions of an open CRN,
while the latter represent the external processes controlling the chemostats.

%%%%%%%%%%%%%%%%%%%%%%%%%%%%%%%%%%%%%%%%%%%%%%%%%%%%%%%%%%%%
%%%%%%%%%%%%%%%%%%%%%%%%%%%%%%%%%%%%%%%%%%%%%%%%%%%%%%%%%%%%
%%%%%%%%%%%%%%%%%%%%%%%%%%%%%%%%%%%%%%%%%%%%%%%%%%%%%%%%%%%%

\bibliography{refs.bib}

\end{document}